\documentclass[reprint,prl,aps,superscriptaddress]{revtex4-2}
\AtBeginDocument{\RenewCommandCopy\qty\SI}
\usepackage{times}
\usepackage{graphicx}
\usepackage{amsmath, braket, amsfonts}
\usepackage{amssymb}
\usepackage{sidecap}
\usepackage{bm, color, ulem}
\usepackage{xcolor}
\usepackage{ragged2e}
\usepackage{wrapfig,lipsum,booktabs}
\usepackage{lineno}
\usepackage{placeins}
\usepackage{tabularx}
\usepackage{makecell}

\usepackage{siunitx}

\newcommand{\be}{\begin{equation}}
\newcommand{\ee}{\end{equation}}

\usepackage{mathtools}
\usepackage{physics}

\DeclareUnicodeCharacter{03BD}{{$\nu$}}

\newcommand{\UCSB}{Department of Physics, University of California at Santa Barbara, Santa Barbara, California 93106, USA}
\newcommand{\nimsTT}{
Research Center for Materials Nanoarchitectonics, National Institute for Materials Science,  1-1 Namiki, Tsukuba 305-0044, Japan
}
\newcommand{\nimsKW}{
Research Center for Electronic and Optical Materials, National Institute for Materials Science, 1-1 Namiki, Tsukuba 305-0044, Japan}

\newcommand{\papertitle}{Visualizing orbital magnetism in electron doped rhombohedral multilayer graphene}

\makeatletter
\def\maketitle{
\@author@finish
\title@column\titleblock@produce
\suppressfloats[t]}
\makeatother

\begin{document}

\title{\papertitle}

\def\UCSB{Physics Department, University of California, Santa Barbara 93106}
\def\StanfordMSE{Department of Materials Science and Engineering, Stanford University, Stanford, CA 94305}
\def\StanfordPhys{Department of Physics, Stanford University, Stanford, CA 94305}
\def\SIMES{Stanford Institute for Materials and Energy Sciences, SLAC National Accelerator Laboratory, Menlo Park, CA 94025}
\def\StanfordAP{Department of Applied Physics, Stanford University, Stanford, CA 94305}

\author{Owen I. Sheekey}
\email{These authors contributed equally} 
\affiliation{\UCSB}
\author{Trevor B. Arp}
\email{These authors contributed equally} 
\affiliation{\UCSB}
\author{Benjamin A. Foutty}
\email{These authors contributed equally} 
\affiliation{\UCSB}
\author{Ruoxi Zhang}
\email{These authors contributed equally} 
\affiliation{\UCSB}
\author{Tixuan Tan}
\affiliation{\StanfordPhys}
\author{Ludwig F. W. Holleis}
\affiliation{\UCSB}
\author{Yi Guo}
\affiliation{\UCSB}
\author{Sandesh~S.~Kalantre}
\affiliation{\SIMES}
\affiliation{\StanfordPhys}
\author{Canxun Zhang}
\affiliation{\UCSB}
\author{Mark Zakharyan}
\affiliation{\UCSB}
\author{David Gong}
\affiliation{\UCSB}
\author{Aidan Keough}
\affiliation{\UCSB}
\author{Youngjoon Choi}
\affiliation{\UCSB}
\author{Ysun Choi}
\affiliation{\UCSB}
\author{Siyuan Xu}
\affiliation{\UCSB}
\author{Tian Xie}
\affiliation{\UCSB}
\author{Ben~Hodder~Alexander}
\affiliation{\SIMES}
\affiliation{\StanfordPhys}
\author{Marisa~Hocking}
\affiliation{\StanfordMSE}
\affiliation{\SIMES}
\author{Qingrui~Cao}
\affiliation{\SIMES}
\author{Martin E. Huber}
\affiliation{Departments of Physics and Electrical Engineering, University of Colorado, Denver, Colorado 80204, USA}
\author{Takashi Taniguchi}
\affiliation{\nimsTT}
\author{Kenji Watanabe}
\affiliation{\nimsKW}
\author{Chenhao Jin}
\affiliation{\UCSB}
\author{\'Etienne Lantagne-Hurtubise}
\affiliation{D\'epartement de physique, Institut quantique \& RQMP, Universit\'e de Sherbrooke, Sherbrooke, Qu\'ebec J1K 2R1, Canada} 
\author{Aaron~Sharpe}
\affiliation{\SIMES}
\affiliation{\StanfordPhys}
\author{Trithep Devakul}
\affiliation{\StanfordPhys}
\author{Andrea F. Young}
\email{andrea@physics.ucsb.edu}
\affiliation{\UCSB}
\date{\today}

\maketitle

\textbf{
Electron doped rhombohedral multilayer graphene at high displacement field features an exceptionally flat band minimum with near-ideal quantum geometry. 
Experiments in this regime observe the formation of a `quarter metal,' in which the electron liquid condenses into a single spin- and valley flavor~\cite{zhou_half-_2021}.  
Remarkably, recent experiments have found a zero resistance state in the same region of the density- and displacement-field-tuned parameter space, attributed to the formation of a chiral superconductor from an orbitally ferromagnetic normal state~\cite{han_signatures_2025}.  
Here, we use nanoSQUID-on-tip magnetometry to map the orbital magnetization of electron-doped rhombohedral graphene devices ranging in thickness between 3 and 15 layers. 
Magnetization within the quarter metal phases peaks at finite density, consistent with concentration of the Berry curvature in a finite-momentum `ring of fire'~\cite{zhang_spontaneous_2011, patri_family_2025}. 
Correlating transport and local magnetometry data in a superconducting tetralayer sample reveals a finite orbital ferromagnetic moment, providing direct evidence of valley polarization in the superconducting ground state.
We further show that widely observed stochastic switching of the resistivity in both metallic and superconducting regimes arises from a density-tuned sign change in the valley-resolved total magnetic moment. 
This leads to the formation of metastable magnetic domains under typical gate control sequences and can also be harnessed for electric-field controlled switching of the magnetization across the entire device.  Finally, high resolution measurements of the magnetization across a superconducting transition allow us to put an upper bound on the `condensation magnetization' of $0.1\mu_B$ per carrier, placing a strong quantitative restriction on theoretical models for ferromagnetic superconductivity\cite{zhu_microscopic_2026}.  
}

In two-dimensional materials possessing low-energy valley degeneracy, interactions can induce spontaneous polarization of the electron system into an itinerant orbital ferromagnet~\cite{sharpe_emergent_2019,serlin_intrinsic_2020,tschirhart_imaging_2021}. 
Unlike its spin counterpart, measurements of the itinerant orbital magnetization can help reveal the quantum geometric properties of the underlying low-energy bands~\cite{xiao_berry_2010, vanderbilt_berry_2018}.
This interplay is particularly rich in crystalline graphene systems, where quantum geometry can be broadly tuned \textit{in situ}.
In monolayer graphene, Berry curvature is singular at the Dirac point, where the dispersion is linear and the bands are degenerate. 
Breaking the sublattice symmetry opens a gap and leads to a finite Berry curvature peaked at the valley points.
In rhombohedral graphene multilayers, applying a perpendicular displacement field $D$ similarly opens a gap near the charge neutrality point. In contrast to the monolayer case, however, the low-energy band edges become extremely flat and the resulting Berry curvature distribution is pushed outward into a finite-momentum ring surrounding each valley, dubbed the `Berry ring of fire'~\cite{zhang_spontaneous_2011, patri_family_2025}.
This highly inhomogeneous and gate-tunable distribution of Berry curvature leads to a dramatic dependence of the orbital magnetization on electron density ($n_e$), displacement field ($D$), and other experimentally tuned parameters---a unique feature of itinerant orbital ferromagnets where the magnetization is determined at the scale of the Fermi wavelength, rather than at the atomic scale.
Recent reports of chiral superconductivity~\cite{han_signatures_2025, qin_extreme_2026, han_evidence_2026} further motivate experimental study of the quantum geometric properties of low-energy bands in rhombohedral multilayer graphene.

More importantly, the reported chiral superconductor itself is expected to host orbital ferromagnetism~\cite{zhu_microscopic_2026}.
However, experimental studies to date have been limited to the electronic transport response, which is dominated by the zero resistance of the superconducting state.
In particular, the anomalous Hall effect---the electronic transport signature of orbital ferromagnetism---cannot be measured in a superconducting state.  
Interpretation of transport anomalies such as stochastic switching of the resistivity in terms of magnetic domain dynamics relies on assumptions about sample homogeneity that cannot be verified by transport measurements alone.  
Local magnetometry addresses some of these limitations by providing a direct measurement of the spatially resolved orbital magnetization~\cite{finkler_scanning_2012}. 
Magnetic imaging is of further interest given the wealth of switching phenomena already known to be present in metallic or insulating orbital magnets, including magnetic reversal actuated by small currents~\cite{sharpe_emergent_2019, serlin_intrinsic_2020,tschirhart_intrinsic_2023}, gate voltages~\cite{polshyn_electrical_2020, han_orbital_2023, choi_superconductivity_2025, zhu_voltage-controlled_2020, bocarsly_electrically_2026, grover_chern_2022}, and circularly polarized light~\cite{holtzmann_optical_2026, persky_optical_2026}.  

\section{Orbital magnetization and the Berry ring of fire} 

\begin{figure}[ht!]
    \centering
\includegraphics[width=\columnwidth]{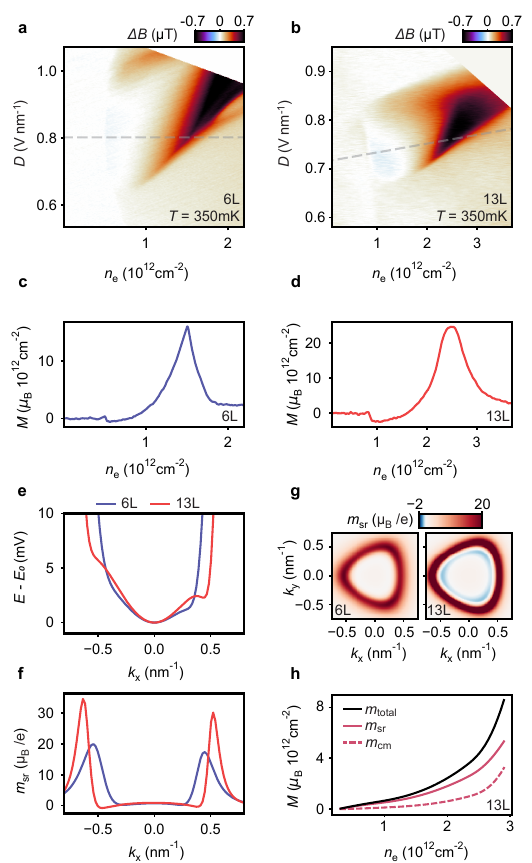} 
\caption{\textbf{Orbital magnetization and `Berry ring of fire' in rhombohedral multilayer graphene}
    \textbf{(a)} Fringe field $\Delta B$ measured at a fixed spatial point above the 6L sample for applied $(B_{||}, B_\perp) = (0\text{ mT}, 12\text{ mT})$ 
    \textbf{(b)} $\Delta B$ for the 13L sample with $(B_{||}, B_\perp) = (0\text{ mT}, 19\text{ mT})$.  
    \textbf{(c)} Calibrated magnetization density $M$ along the grey dashed line in panel a. 
    \textbf{(d)} Calibrated $M$ along the grey dashed line in panel b.   
    \textbf{(e)} Band dispersion near the $K$-point for 6L and 13L rhombohedral graphene for $D = 0.9$ V nm$^{-1}$. 
    \textbf{(g)} Corresponding self-rotation magnetization ($m_{sr}$) for $D = 0.9$ V nm$^{-1}$.  
    \textbf{(f)}  $m_{sr}$ as a function of $k_x$ and $k_y$ for the 6L and 13L for $D = 0.9$ V nm$^{-1}$. 
    \textbf{(h)} Comparison of self-rotation and center-of-mass contributions to the magnetization density, as well as their sum, plotted as a function of density at a constant displacement field of $D = 0.9$ V nm$^{-1}$.
    All data acquired at $T = 350$~mK.
    }
    \label{fig:1}
\end{figure}

Here, we use tilted-tip~\cite{patterson_superconductivity_2025, zhang_imaging_2026} nanoSQUID-on-tip microscopy~\cite{anahory_squid--tip_2020} to directly measure the $n_e$- and $D$-dependent magnetization in a series of rhombohedral multilayer graphene samples with layer numbers $N=3,4,5,6,13$, and $15$, including a tetralayer sample hosting superconductivity within an orbital ferromagnet. Figs.~\ref{fig:1}a--b show the fringe magnetic field measured at a single point above the 6- and 13-layer samples, respectively, taken in the regime of the electron-doped quarter metal~\cite{qin_extreme_2026, guo_flat_2025} (Data for other layer numbers are shown in Fig.~\ref{fig:all_orbital_m}). We measure the magnetization by modulating the gates between the $(n_e,D)$ point of interest and a fixed reference point at several hundred Hz, and recording the resulting magnetic response.
This  measurement gives the difference in static total magnetic fringe field caused by the added electrons, $\Delta B$.  
$\Delta B$ is proportional to the change in total magnetization with a constant of proportionality that depends on the tip position and sample geometry~\cite{patterson_superconductivity_2025}. 
Notably, this constant can be negative (for example, near a device boundary where fringe field flux lines wrap around the device edge), but is positive near the center of a uniform domain of magnetization.
The constant of proportionality can be determined from two dimensional imaging and is constant (for a given position) as long as the magnetic domain structure does not change. 
Figs. 1c-d show fringe field data along the dashed trajectories in panels a and b respectively, with this calibration applied.  
We typically choose a distant reference point where the system is not valley polarized; in this case, $\Delta B$ can be interpreted as an average over many field-cooled configurations.  
Reference points for all data sets are provided in the Methods. 

Measurements in samples of all layer numbers show several striking similarities. 
The fringe fields peak at finite $n_e$ with maximal bulk magnetization of order 10 Bohr magnetons per electron (Fig.~\ref{fig:orbital_moment_estimates}). 
At lower density, in contrast, the magnetization becomes very small. 
Along both trajectories shown in Figs 1c-d the total magnetization drops at low densities within the quarter metal, even appearing to go slightly negative.

\begin{figure*}
    \centering
    \includegraphics[width=\textwidth]{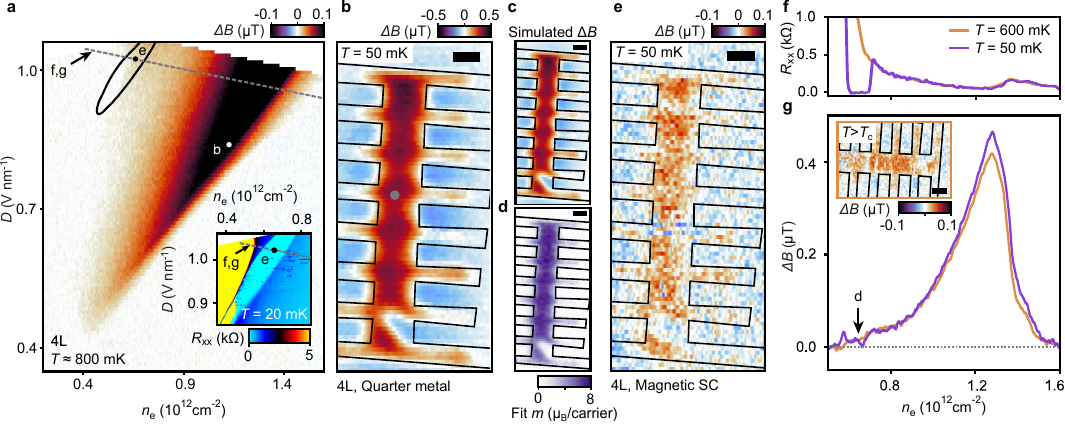} 
    \caption{\textbf{Orbital ferromagnetism in the superconducting state.}
    \textbf{(a)} $\Delta B$ acquired at a fixed spatial point above the 4L sample for applied $(B_\perp, B_\parallel) = (3 \textrm{ mT},50 \textrm{ mT})$ and $T\approx 800$~mK, above superconducting $T_C$. The contour shows the extent of zero resistance in the same sample at 20 mK (inset).  
    \textbf{(b)} Spatial image taken near the peak orbital magnetization in the quarter metal at the $(n_e, D)$ point indicated in panel a.
    \textbf{(c)} Simulated $\Delta B$ arising from (\textbf{d}) spatially distributed magnetization per carrier $m$, which represents a best fit to the data in panel b assuming out-of-plane magnetization.
    \textbf{(e)} Spatial image of $\Delta B$ in the zero resistance state, at the $(n_e, D)$ point indicated in panel a. 
    \textbf{(f)} $R_{xx}$ and
    \textbf{(g)} local fringe field $\Delta B$ measured at $T = 50$~mK and $T = 600$~mK along the trajectory shown by the dashed line in panel a. 
    The inset shows an image with identical conditions to panel e but at $T = 600$~mK, above superconducting $T_c$.  
    For panels b and e-g, the external magnetic field is $(B_{\perp}, B_{\parallel}) = (0.5\text{ mT}, 40\text{ mT})$, and data in panels b and e are measured at $T = 50$ mK. The gray dot in panel b shows the spatial position of measurements of $\Delta B$ in panels a and g. Scale bars are 1 $\mu$m.
    }
    \label{fig:2}
\end{figure*}

These observations can be understood theoretically via the momentum-resolved magnetization arising from wave-packet self-rotations ($m_\text{sr}$)~\cite{xiao_berry_2010}. 
Fig.~\ref{fig:1}e shows the energy dispersion of the conduction band minima for 6- and 13-layer graphene at $D=0.9$ V nm$^{-1}$. 
Fig.~\ref{fig:1}f shows the associated $m_\text{sr}$. 
Both are calculated within a self-consistent Hartree approximation for the $K$ valley (see Supplement). 
At this value of $D$, the conduction band minimum is located at the $K$ point, while $m_\text{sr}$ is nearly zero for small momenta with a large peak at finite momentum. 
Two-dimensional maps of $m_\text{sr}$ near the $K$ point (Fig.~\ref{fig:1}g) show the emergence of a ring of elevated magnetization at finite momenta---the `Berry ring of fire'~\cite{zhang_spontaneous_2011, patri_family_2025}. 
In addition to the wave-packet self-rotation component, the orbital magnetization also includes the so-called center-of-mass contribution, $m_\text{cm}$~\cite{xiao_berry_2010}. 
In Fig.~\ref{fig:1}h we show the total magnetization $M_\text{total}$, $M_\text{sr}$, and $M_\text{cm}$ as a function of density. Both $M_\text{sr}$ and $M_\text{cm}$ show a marked increase at high density. Comparing the model calculations (Fig.~\ref{fig:1}h) with experimental data (Figs.~\ref{fig:1}c--d), we attribute the enhanced magnetization at finite density to carriers populating the finite-momentum single-valley states that host elevated Berry curvature and orbital magnetization.
The subsequent decrease in magnetization observed experimentally arises from a transition out of a fully valley-polarized quarter metal to a spin-polarized `half-metal' state in which the orbital contributions of the two valleys cancel~\cite{zhou_half-_2021}.
As a consequence of this experimentally confirmed theoretical picture, we subsequently refer to the high density peak in magnetization as the `ring of fire.'

Armed with this understanding of the structure of the electron-doped quarter metals, we may contextualize the recent observation of  superconductivity within this phase diagram.  
Fig.~\ref{fig:2}a shows the magnetic fringe field in a superconducting tetralayer sample at an estimated $T\approx 800$~mK, well above the superconducting critical temperature of $T_c\approx 400$~mK (see Fig.~\ref{fig:4LTransport}). 
The black contour indicates the domain where superconductivity emerges at low temperatures (see the inset of Fig. \ref{fig:2}a, which shows resistance measured at $T\approx 20$~mK in the same sample). 
The overall features of this magnetic phase diagram are broadly consistent with those observed at other layer numbers (Figs.~\ref{fig:1} and \ref{fig:all_orbital_m}). 
Superconductivity emerges near the low-$n_e$ boundary of the region where finite magnetization is observed at 800 mK.  
For at least a portion of the superconducting region, the fringe field in the normal state is well above our noise floor, consistent with anomalous Hall measurements that imply a finite orbital magnetization of the normal state (Fig. \ref{fig:4LTransport}). 
Note that transport experiments~\cite{han_signatures_2025,qin_extreme_2026} suggest significant temperature dependence of the low density boundary of the quarter metal which must be considered when comparing the normal and superconducting states at fixed $n_e$ and $D$.

To confirm the magnetic nature of the normal and superconducting states in tetralayer graphene, we image the fringe fields in real space.  Fig. \ref{fig:2}b shows a fringe field map measured near the ring of fire at $T = 50$ mK and an applied field of $(B_\parallel,B_\perp) = (40\text{ mT},0.5\text{ mT})$. 
The measured fringe fields are well fit by simulations of the fringe fields for purely out-of-plane moments localized to the device area and aligned to $B_\perp$ (Methods), shown in Fig. 2c.
Note that due to the applied $B_\parallel\gg B_\perp$, out-of-plane moments are expected to arise only in the presence of easy-axis orbital ferromagnetism.
Fig. \ref{fig:2}e shows a similar $\Delta B$ image with the device tuned into the zero-resistance superconducting state. 
Though the magnitude of the magnetic moment is significantly smaller, the spatial fringe field pattern is qualitatively similar to that near the ring of fire, consistent with an orbital ferromagnetic moment in the superconducting state.
Figs.~\ref{fig:2}f and g show comparisons of the measured resistance and local fringe field for $T$ both above and below the superconducting $T_c$.  Nonzero magnetic fringe field is observed over the extent of the zero resistance state both above and below $T_c \approx400$~mK.
Collectively, these data may be understood as dispositive evidence of orbital magnetization in both the normal state and the superconducting ground state. While the overall scale of $\Delta B$ is similar between $T = 50$~mK and $600$~mK, several features of the detailed structure differ. For example, the magnetization near the ring of fire is slightly temperature dependent, rising at the lowest temperatures. In addition, the $\Delta B$ appears to develop additional structure in the regime of superconductivity.

\section{Capacitively driven domain reversal}

\begin{figure*}
    \centering
\includegraphics[width=\textwidth]{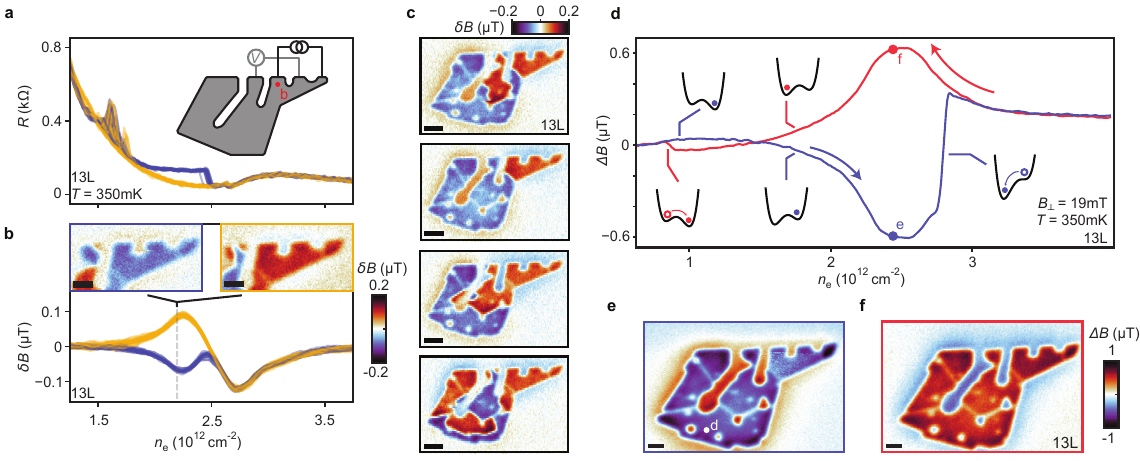} 
\caption{\textbf{Capacitively driven magnetic reversal}
    \textbf{(a)} Resistance measurement from the configuration illustrated in the inset, acquired along the $n_e-D$ trajectory of Fig.~\ref{fig:1}b.  Each line represents a single gate sweep from low to high $n_e$. 
    Curves are colored according to their resistance at $n_e = 2.25 \times 10^{12}$cm$^{-2}$: curves with $R < 100$ ohms are colored yellow, curves with $R > 100$ ohms are colored blue.
    \textbf{(b)} Small signal differential fringe field $\delta B$ measured simultaneously with the transport in panel a at the point indicated in the  inset to panel a. Color coding of traces is consistent with panel a.
    Insets show spatially resolved $\delta B$ for an initialization to $n_e = 2.2\times10^{12}$cm$^{-2}$ for a blue and yellow trajectory, respectively.
    \textbf{(c)} 
    Spatially resolved $\delta B$ over the entire device range. Each image represents a different initialization from $n_e = 0$ into the yellow configuration.
    \textbf{(d)} $\Delta B$ hysteresis along the trajectory in Fig.~\ref{fig:1}b in the absence of a transport current. In the blue (red) curve, the gates are swept such that the system enters the quarter metal from low (high) density.
    \textbf{(e)}, \textbf{(f)} Spatially resolved fringe field maps corresponding to the dots marked on panel d. In panel e, the blue color indicates the entire device is stabilized in the high energy valley.  
    All data acquired at $T = 350$~mK and $(B_{\perp}, B_{\parallel}) = $~(19 mT, 0 mT).
    }
    \label{fig:3}
\end{figure*}

Quantitatively interpreting these signals requires an understanding of real-space structure, particularly the tendency towards formation of nonuniform magnetic domains.  Notably, images taken above and below $T_c$ display apparently  `switchy' behavior, with certain lines appearing to discontinuously jump between positive and negative values (see, e.g., the center of Fig. \ref{fig:2}d). 
Stochastic changes are also apparent as line-by-line jumps in the transport measurements in the inset of Fig. \ref{fig:2}a (see also Fig.~\ref{fig:4LTransport}). 
These features appear even within the superconducting state, manifesting as a finite resistance for certain measurements of $R_{xx}$ within a domain largely characterized by $R_{xx}=0$.
Such stochastic switching is ubiquitous in published transport data in rhombohedral quarter metals~\cite{zhou_superconductivity_2021}, and is commonly attributed to the formation of magnetic domain walls, which may be dissipative even in a superconducting state~\cite{han_signatures_2025}. 
Fig.~\ref{fig:3}a shows resistance measured in the configuration shown, for a 13 layer device acquired by sweeping $n_e$ from low to high along the trajectory shown in Fig. \ref{fig:1}b.
The stochastic switching manifests via the clustering of resistance traces into two classes which we color code yellow and blue, respectively. Blue curves appear to correspond to a metastable configuration which relaxes suddenly near $n_e\approx 2.5 \times 10^{12}$~cm$^{-2}$ to the smoother trajectory defined by the yellow curves.  
Fig. \ref{fig:3}b shows the fringe magnetic field measured simultaneously with the transport data in Fig. \ref{fig:3}a at a fixed spatial point (see inset to Fig. \ref{fig:3}a), with the curves color coded by their resistance value.     
We plot the differential magnetometry signal $\delta B$, defined as the magnetic fringe field response to a small finite frequency sinusoidal modulation of the bottom gate with amplitude $\delta V^{\text{RMS}}_{bg} = 20$ mV (corresponding to a simultaneous modulation of $\delta n_e\approx 0.026 \times 10^{12}$ cm$^{-2}$ and $\delta D\approx 2.91$mV nm$^{-1}$).
While $\delta B$ is not directly proportional to the total magnetization, it may still serve to determine the underlying magnetic state, as time reversed states will show equal and opposite $\partial m/\partial V_{bg}$. 
Unlike $\Delta B$, which averages the result of repeated gate initializations, $\delta B$ allows us to image a single stable domain configuration.
The simultaneous clustering of resistance and $\delta B$ curves indicates that resistance switching arises from differing local magnetic response.

To verify the origin of this effect in terms of real space domains, we sweep $n_e$ to $n_e=2.3\times 10^{12}$~cm$^{-2}$ and then perform spatial imaging of $\delta B$ at fixed $n_e$.  The two insets to Fig.~\ref{fig:3}b show $\delta B$ for representative examples in the two classes of resistance curves. Evidently, blue curves correspond to a metastable configuration in which a several-micron domain near the transport contact is oriented opposite the applied magnetic field of $B_\perp = 19$~mT, while the yellow curves correspond to a state where the same region is polarized in the direction of the applied field. Of course, while the transport response is most sensitive to magnetic inhomogeneity near the contacts, more complex domain structures are also observed.  Fig.~\ref{fig:3}c shows images of the entire device following the same sequence described above but entirely within the yellow class of curves.  Different patterns of reversed domains are clearly visible for the same initialization sequence and effectively identical $R$ measurements.  Such inhomogeneity must be accounted for in the interpretation of transport measurements.  

\begin{figure*}
    \centering
    \includegraphics[width=\textwidth]{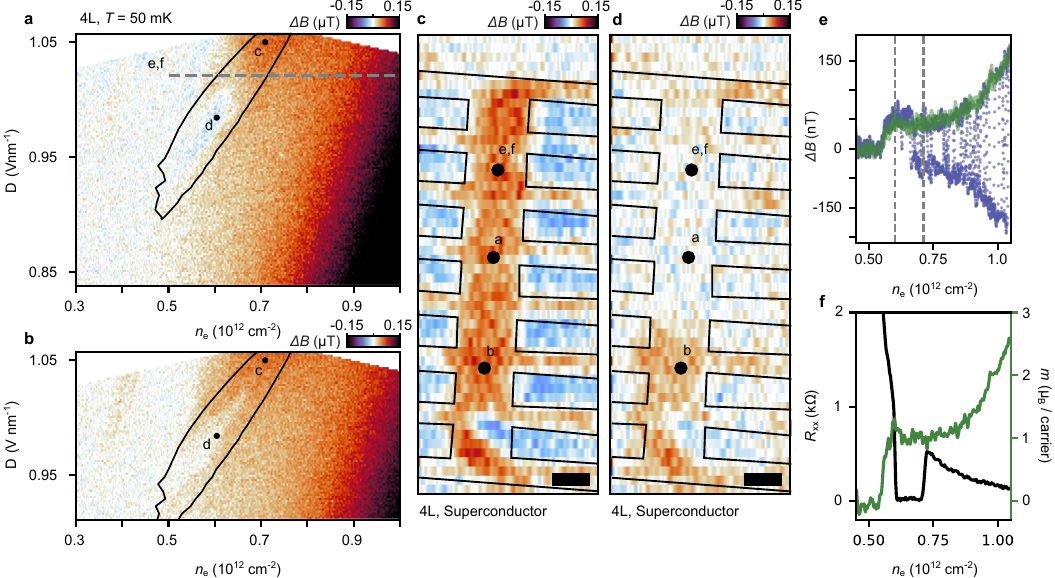} 
    \caption{\textbf{Magnetic domains and condensation magnetization in the  superconducting state}
    \textbf{(a)} $\Delta B$ measured as a function of $n_e$ and $D$ in device 4L. The solid line shows the contour of superconductivity determined from transport measurements. 
    \textbf{(b)} The same as panel a at a different location over the device.
    \textbf{(c)} Spatial image of $\Delta B$ acquired at $n_e=0.71 \times 10^{12}\text{ cm}^{-2}$ and $D=1.05 \text{ V nm}^{-1}$ with reference point $n_e=3.96 \times 10^{12}\text{ cm}^{-2}$ and $D= 0.76\text{ V nm}^{-1}$. 
    \textbf{(d)} Spatial image of $\Delta B$ acquired at $n_e=0.60 \times 10^{12}\text{ cm}^{-2}$ and $D=0.98 \text{ V nm}^{-1}$. 
    The spatial positions corresponding to measurements shown in panels a, b, e, and f are marked. Panels a-d are measured using a large 
    $\delta V_{BG}$, initializing the quarter metal through the ring of fire.
    \textbf{(e)} $\Delta B$ as a function of $n_e$ at $D = 1.02$ V/nm. The data are taken under two reference conditions. In blue, the differential $\Delta B$ is measured relative to a nonmagnetic state at $(n_e,D) = (0.24\times 10^{12}\textrm{cm}^{-2}, 1.02\textrm{ V nm}^{-1})$. In green, $\Delta B$ is measured with a large $\delta V_{BG}$, similar to panels a-d. 
    \textbf{(f)} $R_{xx}$ (left axis) as a function of $n_e$ for $D = 1.02$ V nm$^{-1}$ compared with the magnetization per carrier $m$ from panel \textbf{e}. All data in the figure are measured at $(B_\perp, B_\parallel) = (0.5 \text{ mT}, 40\text{ mT})$ and $T = 50$ mK. Scale bars are 1$\mu$m. 
    }
    \label{fig:4}
\end{figure*}

We attribute the stabilization of reversed magnetic domains to the gate-controlled change in sign of the single valley total magnetization~\cite{zhu_voltage-controlled_2020, das_unconventional_2023}.  This phenomenon, unique to orbital magnets, has attracted attention 
~\cite{polshyn_electrical_2020, han_orbital_2023, choi_superconductivity_2025, bocarsly_electrically_2026} because it allows magnetization to be reversed by purely capacitive means. 
To illustrate this effect, in Fig.~\ref{fig:3}d we show the fringe field $\Delta B$ for reference points 
at either end of the same $n_e-D$ trajectory as in Figs.~\ref{fig:3}a-b.
The two sweep directions show nearly opposite total magnetic response: for the trajectory originating at high density we observe a large magnetization aligned with the applied magnetic field at high $n_e$ which becomes small and negative at low $n_e$.
Conversely, for the trajectory originating at low $n_e$, we observe a large and negative magnetization at high $n_e$, which is small and positive at low $n_e$.  
To ensure our local measurements are reflective of the global behavior of the device, we perform spatial imaging at the maximum $|B|$ points along these curves, shown in Figs.~\ref{fig:3}e and f for rising- and falling $n_e$ traces, respectively.  
The fringe fields in Fig.~\ref{fig:3}(e) and (f) are nearly uniformly reversed, indicating that the two states have equal and opposite magnetization.
As shown in Figs.~\ref{fig:5L_switching}, Fig.~\ref{fig:6L_switching}, and Fig.~\ref{fig:alt_13L_switching}, qualitatively similar behavior is also observed in 5-, 6-, and a second 13-layer device.  We note that we were able to stabilize the uniform reversed domain  of Fig. 3e only in the absence of an applied transport current, the presence of which tends to favor mixed configurations of the type of Fig. 3c.  

The essential mechanism for this reversal is illustrated in the insets to Fig. \ref{fig:3}d, which depict a schematic `double well' potential in which each minimum represents macroscopic occupation of the $K$ or $K'$ valley.  In an applied field, the degeneracy between the two time-reversed states is lifted, so that one valley is energetically favored.  The observation of gate tuned magnetic switching can be understood if the system stabilizes in the low energy valley upon entry to the quarter metal but relaxation between valleys is suppressed within the quarter metal itself.  For the blue curve, the system enters the low-energy valley (say, $K$) at low $n_e$, and remains in this valley even as $M_K$, the total magnetization of the occupied orbitals, changes sign.  Fig. \ref{fig:3}e is a spatial image of this state with magnetization opposite the applied field.  On further increase of $n_e$, however, relaxation to the $K'$ valley occurs, manifesting as a sudden reversal of $\Delta B$ near $n_e=2.75 \times 10^{12}$ cm$^{-2}$.  For the red curves, the system enters lower energy $K'$ valley at the high density boundary of the quarter metal, and remains there even as $M_{K'}$ reverses sign before relaxing to the $K$ valley at $n_e\approx 0.9\times 10^{12}$ cm$^{-2}$.  Electrically  controllable magnetic reversal was first proposed~\cite{zhu_voltage-controlled_2020} and observed via transport measurements~\cite{polshyn_electrical_2020} in moire Chern insulators.  Our results here represent the first real-space imaging of this unique orbital magnetic phenomenon, and demonstrate that it is a generic feature of metallic orbital ferromagnets in rhombohedral graphene across a wide range of layer numbers, and must be accounted for in interpreting transport measurements whenever the ground state features valley imbalance. 

\section{Domains and `condensation magnetization' in the superconducting state}

We now return to the magnetic superconducting state in Device 4L. 
In Fig. \ref{fig:4}a-b, we present measurements of $\Delta B$ as a function of $n_e$ and $D$ measured at two distinct locations above the device.  
In both panels, we take the reference point to be within the spin polarized half metal, entering the quarter metal through the ring of fire in order to favor single domain structure (Figs.~\ref{fig:3},\ref{fig:5L_switching}  \ref{fig:6L_switching}, and \ref{fig:alt_13L_switching}). 
Both panels show sizable magnetic signal throughout much of the superconducting region, outlined in black. 
Notably, we find no magnetic signature associated with the onset of  superconductivity.
This is in marked contrast to a recent experiment performed under nearly identical conditions on a different, nonmagnetic rhombohedral graphene superconductor in which a clear diamagnetic response was observed precisely coinciding with the onset of superconductivity\cite{zhang_imaging_2026}. This suggests that any Meissner response at the applied field of $B_\perp = 500~\mu$T is considerably smaller than our noise floor of approximately 5 nT.  
However, comparing the two phase diagrams, we do observe a distinct region of diminished or reversed magnetic response within the superconducting state, which appears to vary in strength and extent at the two spatial locations.  Fig. \ref{fig:4}c shows a spatial image of $\Delta B$ in the high-$D$ part of the superconductor, far from this feature. The map is consistent with a single spatially uniform domain of magnetization as in Fig. \ref{fig:2}.  As shown in Fig. \ref{fig:4}d, however, imaging in the vicinity of the reduced magnetic feature shows a sharply inhomogeneous response. 
Specifically, only the lower end of the device appears to support a stable magnetization, while the upper half of the device appears to have little or no net magnetization.  We interpret this contrast as indicating instability of the magnetization to reversal under our gate pulsing protocol. As discussed for Device 13L in Fig. \ref{fig:dynamics}, if repeated `field cooling' along the gate modulation trajectory does not strongly favor one polarization over another, a zero signal can result as positive and negatively magnetized domains are averaged over many initialization cycles.    

This hypothesis is further substantiated in Fig. \ref{fig:4}e, which compares $\Delta B$ at a single spatial point for different choices of reference point.  When the reference point is chosen as in panels a-d, a smooth curve results, as expected for a consistent domain configuration.  However, when the reference point is chosen at  $(n_e,D) = (0.24\times 10^{12}\textrm{cm}^{-2}, 1.02\textrm{V/nm})$,  we observe bistable behavior---including within the superconducting state. The origin of the bistability itself is less clear, as each point in panel e represents approximately 10 cycles in which the density is tuned from the reference to the measurement point; if each cycle represents an independent field cooling through the magnetic transition, one expects a range of possible values.  Rather, the bistability may arise if transient heating from the gate modulation is sufficient to cause thermal domain reversal rates in the 1 Hz range, leading to the observed $\mathcal{O}(10)$ reversals on the few minute timescale of our measurements. We note that similar behavior is observed in spatial imaging of the device above $T_c$ under similar gate configurations (Fig. \ref{fig:4L_aboveTc_switching}), and such switching appears to be suppressed by an increased $B_\perp$. Regardless of the precise mechanism, however, the observation of bistable behavior of $\Delta B$ in the superconducting state provides further evidence for the existence of two nearly degenerate magnetization states, as expected for a valley polarized orbital ferromagnet.  
This detailed understanding of the domain dynamics allows us to study the intrinsic magnetization of the superconducting state more quantitatively. Fig. \ref{fig:4}f compares the total magnetization per carrier, $m$, with the sample resistance across the superconducting state.  This measurement reveals that the magnetization is indistinguishable from 1 $\mu_B/\textrm{carrier}$ across the entire superconducting regime. This finding is consistent with a \textit{vanishing} orbital part of the magnetization despite full valley polarization, and the observed $m$ arises from locking of the spin to the valley due to intrinsic spin-orbit coupling\cite{arp_intervalley_2024}.  More importantly, we find no detectable signature of the superconductor itself, with $M$ smoothly evolving across the boundary between the superconductor and the adjacent valley imbalanced normal metal.  For a superconductor in which the Cooper pair condensate carries net angular momentum relative to the normal state, a magnetic signature of the superconducting transition is in principle expected.  Our measurement bounds this effect at no more than $0.1\mu_B/\textrm{carrier}$, providing a quantitative experimental constraint on theoretical models of valley polarized superconductors\cite{zhu_microscopic_2026}.

As a final experimental finding, we note that in the 6-layer sample studied in Fig.~\ref{fig:1}a, we observed no signatures of superconductivity in the electron-doped quarter metal, contrasting with literature reports\cite{qin_extreme_2026, han_evidence_2026}.  Measurements of the same device show several superconducting states at hole doping with comparatively low $T_c\approx 100\text{ mK}$, suggesting the absence is an intrinsic feature of the device rather than an artifact of elevated electron temperature (see Fig.~\ref{fig:hexalayer_transport}).  Our local magnetometry measurements reveal a novel source of spatial inhomogeneity in this particular device, in which only a fraction of the device ever becomes magnetic in the low $n_e$, high $D$ regime where superconductivity was reported (see Extended Data Fig.~\ref{fig:hexalayer_inhomoheneity}).  While the origin of this inhomogeneity cannot be definitively established, we speculate that strain may play a key role, selecting ground states by their nematic susceptibility and influencing the phase diagram (see Supplementary information for further discussion).  Whether strain favors or disfavors the orbital ferromagnets hosting superconductivity could be established in the future by relating local strain to the ground state magnetization.

\section{Discussion}

Chiral superconductivity is conventionally defined~\cite{kallin_chiral_2016} as the momentum space winding of the complex gap function.  
This leads to an orbital magnetization associated with the superconducting state itself.  
Emergence of a chiral superconductor from a non-magnetic normal state should thus appear as a finite orbital magnetic polarization of the superconducting state, and the observation of such a polarization onset at $T_c$ would constitute strong evidence for chiral superconducting order parameters~\cite{robbins_theory_2020}. 
For the case of a single, simply-connected Fermi surface, the theoretical expectation is that any superconducting state with finite orbital magnetization should also be chiral in the sense of a winding order parameter.
The rhombohedral magnetic superconductor adds two key features to the situation considered in the theoretical literature developed for most other experimental systems.  
First, the normal state is already an orbital ferromagnet.   Second, the low-density phase diagrams of all rhombohedral graphene multilayers feature numerous Lifshitz transitions separating phases with, at times, complex, multi-pocket fermiology.  Original reports of magnetic superconductivity did not experimentally determine the fermiology, which was assumed to be similar to that of a nearby single Fermi surface valley polarized quarter metal.  More recent experiments on the same tetralayer device studied here \cite{kalantre_candidate_2026} suggest a more complex fermiology in the superconducting parent state.  

These observations motivated a more inclusive definition of chiral superconductivity\cite{geier_chiral_2025} as any superconductor coexisting with finite orbital magnetization, regardless of the detailed structure of the gap function.
Our observation of finite orbital magnetization in both the superconducting and normal states is dispositive evidence of chiral superconductivity under this definition.  However, it has no bearing on whether the superconducting order parameter has nontrivial winding around the Fermi surface(s) or whether the Cooper pairs themselves pair with finite relative angular momentum. 
Indeed, the presence of magnetization in the normal state makes assessing the magnetization of the Cooper pair condensate relative to the normal state more difficult. Specifically, the onset of superconductivity is expected to change the magnetization due not only to this `condensate magnetization' but also to the change in occupation of the Bloch states near the Fermi level.  Recent theoretical work suggests that this interplay may be very delicate\cite{zhu_microscopic_2026}, leading to either an increase or decrease in $m$ upon condensation.  The lack of observed change in $m$---determined here to the $0.1\mu_B$ per carrier level---provides a key experimental constraint on theories of the magnetic superconductor observed in rhombohedral graphene. 

\textit{Note added}: During preparation of this manuscript, we became aware of a work also reporting imaging of orbital ferromagnetism in a rhombohedral graphene superconductor~\cite{dutta_reconfigurable_2026}.

\bibliographystyle{science}
\bibliography{references,other_references}

\section*{Acknowledgments} 
The authors would like to acknowledge discussions with Erez Berg, Mathieu Haguier, Julian May-Mann and Yaar Vituri.
Work in the Young lab at UCSB was primarily supported by a MURI project of the Air Force Office of Scientific Research under grant number FA9550-25-1-0287.  
Additional support provided by the Gordon and Betty Moore Foundation's Experimental Physics Investigator award GBMF13801 and the W. M. Keck Foundation under award SB190132 for the development of SQUID microscopy techniques. 
T.A. and O.S. acknowledge direct support by the National Science Foundation through Enabling Quantum Leap: Convergent Accelerated Discovery Foundries for Quantum Materials Science, Engineering and Information (Q-AMASE-i) award number DMR-1906325; the work also made use of shared equipment sponsored under this award. 
K.W. and T.T. acknowledge support from the JSPS KAKENHI (Grant Numbers 21H05233 and 23H02052) and World Premier International Research Center Initiative (WPI), MEXT, Japan. E.L.-H is grateful for support from the Natural Sciences and Engineering Council of Canada (NSERC), grants RGPIN-2025-06136 and ALLRP-613856-25, the Regroupement québécois sur les matériaux de pointe (RQMP, \href{https://doi.org/10.69777/309032}{https://doi.org/10.69777/309032}), and start-up funds from the Faculté des Sciences at Université de Sherbrooke.
T.D. and T.T. acknowledge support from the Air Force Office of Scientific Research under award number FA9550-25-1-0343.
Part of this work was performed at the Stanford Nano Shared Facilities (SNSF), supported by the National Science Foundation under Award No. ECCS-2026822, as well as in the nano@Stanford labs, which are supported by the National Science Foundation as part of the National Nanotechnology Coordinated Infrastructure under Award No.ECCS-1542152. M.H. acknowledges partial support from the U.S. Department of Defense through the Graduate Fellowship in STEM Diversity program.
Fabrication and analysis performed by S.K., M.H, Q.C, B.A, and A.S. were supported by the US Department of Energy, Office of Science, Basic Energy Sciences, Materials Sciences and Engineering Division, under Contract DE-AC02-76SF00515. B.H.A. was supported by the Gordon and Betty Moore Foundation under Grant No. GBMF9460.

\section*{Author contributions}
OIS, TBA, BAF, RZ, and CZ performed the nanoSQUID-on-tip measurements. 
TT and TD completed the theoretical calculations.
TT, ELH, and TD provided theoretical support.
OIS, TBA, BAF, RZ, LFWH, YG, and CZ contributed to analysis and understanding. 
LFWH, BAF completed low temperature transport measurements. 
SX and TX fabricated the 3 layer sample supervised by CJ.
SSK, BHA, MH, and QC characterized and fabricated the 4 layer sample, supervised by AS.
YSC and YJC fabricated the 5 layer sample.
LFWH fabricated the 6 layer sample.
YG fabricated the primary 13 layer sample, LFWH fabricated the secondary 13 layer sample. 
AFY supervised fabrication of 5, 6 and 13 layer samples.
BAF, CZ, AK, and DG fabricated nanoSQUIDs.
MZ developed custom electronics used for arbitrary wave generation.
TT and KW provided hBN crystals. 
MEH provided SQUID array amplifiers for nSOT readout. 
OIS, TBA, BAF, RZ, LFWH, and AFY analyzed the data. 
OIS, TBA, and AFY wrote the paper with inputs from all authors. 
AFY supervised and conceived the project.

\section*{Competing interests}
The authors declare no competing interests.

\section*{Data availability}
All supporting data for this paper and other findings of this study are available from the corresponding authors upon request. 

\section*{Code availability}
All numerical codes for this paper and other findings of this study are available from the corresponding authors upon request. 


\section*{Methods}

\subsection*{Sample fabrication}

In this work we present data from seven different devices which we label based on their layer number from three to fifteen. In the following we describe the fabrication specifics of each device.\\

\textbf{Device 3L}: the device is stacked in two steps. The bottom half consisting of graphite gate and hBN is stacked with  poly carbonate stacking film which is dissolved in Chloroform.
Subsequently the sample is heated to 375~$^{\circ}$C in vacuum for several hours to remove polymer residues. 
The top half is then assembled with a flat PDMS stamp at pickup angles $< 1$ degree between the PDMS stamp and the SiO$_2$ substrate.
One region of the dual gated rhombohedral trilayer region is proximitized to WSe$_2$. 
Another region close to the contacts only sees hBN substrates on both the top and the bottom, above which the phase diagram in Fig.~\ref{fig:all_orbital_m} was measured.
This device was previously studied in Ref.~\cite{patterson_superconductivity_2025}.

\textbf{Device 4L}:
Rhombohedral regions in tetra-layer graphene flakes were identified using Raman spectroscopy alongside amplitude-modulated Kelvin probe force microscopy. The crystallographic orientation was discerned using atomically-resolved torsional force microscopy (TFM)~\cite{pendharkar_torsional_2024}.  The rhombohedral regions were subsequently isolated from Bernal regions via anodic oxidation. The device is stacked in three steps. The bottom piece consisting of graphite gate and hBN is stacked with  poly carbonate stacking film which is dissolved in chloroform and cleaned with contact mode AFM. Then, hBN/R4G was assembled stacking along the zigzag direction, and then released onto the bottom piece, which is again cleaned using chloroform and contact mode AFM.  Finally, an hBN flake/top graphite gate are picked up and released on the completed heterostructure. The film was dissolved using chloroform, but no subsequent AFM cleaning was performed. Tapping AFM was used to identify a bubble-free region of the completed heterostructure, which was then etched into a Hall bar geometry and contacted using standard e-beam lithography techniques. 

\textbf{Device 5L}: the device is stacked in two steps. 
First, a graphite bottom gate on hBN is picked up with a poly(bisphenol A carbonate) film, and bilayer WSe$_2$ is picked up subsequently. 
The stack is then flipped using a gold coated PDMS stamp following the technique developed in ref.~\cite{kim_imaging_2023}.
This leaves a pristine WSe$_2$/hBN surface serving as the bottom half for the top half of the stack to be placed on.
The top half consisting of pickup hBN, top gate, hBN, and rhombohedral five layer graphene is stacked sequentially with the low pressure pickup technique developed in Ref.~\cite{holleis_cryogenic_2026} and then dropped onto the bottom half.

\textbf{Device 6L, 13L-2, and 15L}: rhombohedral multilayer graphene for these devices was exfoliated via cryogenic shock exfoliation and characterized via scanning Microwave Impedance Microscopy (sMIM)~\cite{holleis_nanoscale_2025}, with the full fabrication procedure following Ref~\cite{holleis_cryogenic_2026}.
Device 13L-2 was previously studied in the main text of that article.

\textbf{Device 13L}: rhombohedral graphene for this device was prepared via standard mechanical exfoliation. The rhombohedral domains were identified using Raman spectroscopy~\cite{lui_imaging_2011,zhang_raman_2016}, infrared imaging~\cite{lu_extended_2025,feng_rapid_2025}, and sMIM~\cite{holleis_nanoscale_2025}. Unique to this device, and for the purposes of symmetric device geometry, special care was taken in dielectric preparation. The top and bottom hBN dielectrics were prepared by cleaving a single, thin flake into two identical halves using a (P-50 PtIr) STM tip on the transfer station. All subsequent stacking and cleaning procedures followed Ref.~\cite{guo_flat_2025}.


For devices 3L, 5L, 6L, 13L, 13L-2, and 15L, we first deposit the contacts to the rhombohedral layers, and then perform a device defining etch.

\begin{table}[h]
\centering
\begin{tabularx}{\columnwidth}{c|c|c|c}
\textbf{Sample} & \makecell{\textbf{Top/Bottom hBN} \\ \textbf{thickness (nm)}} & \textbf{WSe$_2$} & \textbf{(Extended data) figures} \\
\hline
 3L & 8/10 & yes/no & \ref{fig:all_orbital_m}\\
 4L & 17/25 & no & \ref{fig:2},\ref{fig:all_orbital_m},\ref{fig:4LTransport}\\
 5L & 15/17 & yes & \ref{fig:all_orbital_m},\ref{fig:5L_switching}\\
 6L & 18/19 & no & \makecell{\ref{fig:1},\ref{fig:4},\ref{fig:all_orbital_m},\\ \ref{fig:6L_switching},\ref{fig:hexalayer_transport}}\\
 13L & 12/12 & no & \makecell{\ref{fig:1},\ref{fig:3},\ref{fig:all_orbital_m},\\ \ref{fig:orbital_moment_estimates}}\\
 13L-2 & 25/25 & no & \ref{fig:alt_13L_switching}\\
 15L & 30/15 & no & \ref{fig:all_orbital_m}\\
\end{tabularx}
\caption{Device parameters}
\label{table:samples}
\end{table}

\subsection*{Details of nanoSQUID on tip measurements}
The nanoSQUID on Tip (nSOT) measurements were performed using indium SQUIDs fabricated on the end of pulled quartz pipettes using the self-aligned fabrication technique~\cite{anahory_squid--tip_2020}. 
Multiple nSOTs were used across all data, with typical diameters on the order of 300~nm and sensitivity on the order of 3~nT/$\sqrt{\textrm{Hz}}$. A large diameter was selected to ensure our flux biasing field was small out-of-plane. Furthermore the nSOT tips were angled 12.5$^{\circ}$ off of vertical to allow the option of flux biasing by applying large in-plane fields and low out-of-plane fields, which can reduce the extraneous signal from gate Landau levels and spin magnetism~\cite{patterson_superconductivity_2025}. 
The nSOT signal was read out by measuring the current through the tip with a series SQUID array amplifier in feedback mode~\cite{huber_dc_2001}.
The nSOT was mechanically controlled using a piezoelectrically excited quartz tuning fork in a phase-locked loop positioned over the sample with an Attocube ANSxyz100std/LT xyz-Scanner to control the tip $\approx 200$~nm above the RMG layer.
Measurements were typically performed using a square-wave or AWG modulation, described below, but data in Fig.~\ref{fig:3}b,c were gathered with differential magnetometry, $\delta B$, by applying a small a.c. modulation to the bottom gate and locking into the nSOT with a Stanford Research Systems SR830, as described in Ref.~\cite{arp_intervalley_2024}.
Below, we detail the precise experimental conditions for each of the datasets in the main text.
\begin{itemize}
    \item Fig.~\ref{fig:1}(a),(c): Device: 6L, $(B_{\perp}, B_{\parallel}) = $~(12 mT, 0 mT), 260~nm diameter nSOT 150~nm above surface (180~nm above RMG), Measurement frequency: 1.743 kHz, nSOT sensitivity: $\approx$ 3~nT/$\sqrt{\textrm{Hz}}$, $T = $~350~mK. Square wave reference point: $(n_e, D) = (2.22 \times 10^{12}\text{cm}^{-2},0.8\text{V/nm})$
    \item Fig.~\ref{fig:1}(b),(d): Device: 13L, $(B_{\perp}, B_{\parallel}) = $~(19 mT, 0 mT), 240~nm diameter nSOT 150~nm above top surface (200~nm above RMG), Measurement frequency: 1.77 kHz, nSOT sensitivity: $\approx$ 3~nT/$\sqrt{\textrm{Hz}}$, $T = $~350~mK. Square wave reference point: $(n_e, D) = (3.97 \times 10^{12}\text{cm}^{-2},0.73\text{V/nm})$
    
    \item Fig.~\ref{fig:2}(a): Device: 4L, $(B_{\perp}, B_{\parallel}) = $~(3mT, -50 mT), 350~nm diameter nSOT 300~nm above surface (350~nm above RMG), Measurement frequency: 381 Hz, Square wave reference: $(n_e, D) = (1.41 \times 10^{12}\text{cm}^{-2},0.51\text{V/nm})$, nSOT sensitivity: $\approx$ 2~nT/$\sqrt{\textrm{Hz}}$, the temperature at the mixing chamber was $50$ mK, but He exchange gas in the system resulted in an elevated temperature due to heating from the nSOT~\cite{zhang_imaging_2026}, significantly above the superconducting $T_c$ which onsets around $500$ mK. From transport comparisons, the effective electron temperature is likely around $T\approx 800$ mK in this condition.

    \item Fig.~\ref{fig:2}(b,e,g): Device: 4L, $(B_{\perp}, B_{\parallel}) = $~(0.5 mT, 40 mT), 350~nm diameter nSOT 300~nm above surface (350~nm above RMG), Measurement frequency: 167 Hz, nSOT sensitivity: $\approx$ 4~nT/$\sqrt{\textrm{Hz}}$, $T = 50$ mK (b,e,g) or $T = 600$ mK (g). For these measurements, He exchange gas was removed from the microscope, and there is no significant heating from the nSOT. Fig.~\ref{fig:2}(b) $(n_e, D) = (1.21 \times 10^{12}\text{cm}^{-2},0.87\text{V/nm})$, square wave reference $(n_e, D) = (1.66 \times 10^{12}\text{cm}^{-2},0.83\text{V/nm})$ ($\delta V_{BG} = +0.7$ V). Fig.~\ref{fig:2}(e) and Fig.~\ref{fig:2}(g) inset, $(n_e, D) = (0.64 \times 10^{12}\text{cm}^{-2},1.02\text{V/nm})$, square wave reference $(n_e, D) = (0.54 \times 10^{12}\text{cm}^{-2},1.02\text{V/nm})$ ($\delta V_{BG} = -0.15$ V). Fig. ~\ref{fig:2}(g) linetrace, square wave reference $(n_e,D) = (3.89 \times 10^{12}\text{cm}^{-2},0.72\text{V/nm})$.

    \item Fig.~\ref{fig:3}(b): Device: 13L, $(B_{\perp}, B_{\parallel}) = $~(19 mT, 0 mT), 240~nm diameter nSOT 150~nm above surface (200~nm above RMG), Measurement frequency: 1.77 kHz, nSOT sensitivity: $\approx$ 3~nT/$\sqrt{\textrm{Hz}}$, a.c. gate voltage: $\delta V_{bg} = 20$mV, $T = $~350~mK. 
    \item Fig.~\ref{fig:3}(b insets),(c): Device under measurement: 13L, $(B_{\perp}, B_{\parallel}) = $~(19 mT, 0 mT), 240~nm diameter nSOT 300~nm above surface (350~nm above RMG), Measurement frequency: 1.77 kHz, nSOT sensitivity: $\approx$ 3~nT/$\sqrt{\textrm{Hz}}$, a.c. gate voltage: $\delta V_{bg} = 20$mV, $T = $~350~mK. 
    \item Fig.~\ref{fig:3}(d): Device: 13L, $(B_{\perp}, B_{\parallel}) = $~(19 mT, 0 mT), 240~nm diameter nSOT 150~nm above surface (200~nm above RMG), Measurement frequency: 1.77 kHz, nSOT sensitivity: $\approx$ 3~nT/$\sqrt{\textrm{Hz}}$, $T = $~350~mK. Increasing density sweep (blue) referenced to $(n_e,D) = (0 \times 10^{12}\text{cm}^{-2},0.7\text{V/nm})$. Decreasing density sweep (red) referenced to $(n_e,D) = (3.97 \times 10^{12}\text{cm}^{-2},0.79\text{V/nm})$.
    \item Fig.~\ref{fig:3}(e),(f): Device: 13L, $(B_{\perp}, B_{\parallel}) = $~(19 mT, 0 mT), 240~nm diameter nSOT 300~nm above surface (350~nm above RMG), Measurement frequency: 1.77 kHz, nSOT sensitivity: $\approx$ 3~nT/$\sqrt{\textrm{Hz}}$, $T = $~350~mK. Both (e) and (f) referenced to $(n_e,D) = (3.97 \times 10^{12}\text{cm}^{-2},0.79\text{V/nm})$. In (f) the opposite valley is achieved by modulating the gates to the `initialization value' of $(n_e,D) = (0.35 \times 10^{12}\text{cm}^{-2},0.98\text{V/nm})$ over a period of 96$\mu$s before setting them to the magnetic point. This protocol ensures that both (e) and (f) have the same reference point but are initialized in different ways.
    
    \item Fig.~\ref{fig:4}(a-f): Device: 4L, $(B_{\perp}, B_{\parallel}) = $~(0.5 mT, 40 mT), 350~nm diameter nSOT 300~nm above surface (350~nm above RMG), Measurement frequency: 167 Hz, nSOT sensitivity: $\approx$ 4~nT/$\sqrt{\textrm{Hz}}$, $T = 50$ mK. Fig.~\ref{fig:4}(a-b), square wave reference is a constant $\delta V_{BG} = +2 V$, corresponding to $(\delta n_e,\delta D) = (1.30\times 10^{12}\text{cm}^{-2},-0.12 \text{V/nm})$. The reference for both images is entirely contained in the valley-balanced half metal with zero magnetic fringe field due to the in-plane applied magnetic field. Fig.~\ref{fig:4}(c), $(n_e, D) = (0.71 \times 10^{12}\text{cm}^{-2},1.05\text{V/nm})$, square wave reference $(n_e, D) = (3.96 \times 10^{12}\text{cm}^{-2},0.76\text{V/nm})$ ($\delta V_{BG} = 5 V$). Fig.~\ref{fig:4}(d), $(n_e, D) = (0.60 \times 10^{12}\text{cm}^{-2},0.98\text{V/nm})$, square wave reference $(n_e, D) = (3.85 \times 10^{12}\text{cm}^{-2},0.69\text{V/nm})$ ($\delta V_{BG} = 5 V$). Fig.~\ref{fig:4}e, Square wave reference $(n_e, D) = (0.24 \times 10^{12}\text{cm}^{-2},1.02\text{V/nm})$.
    
\end{itemize}

\subsection*{Transport measurements}

Transport measurements on Device 4L are all taken in a standard four-terminal geometry between 13 and 27 Hz, with a.c. current limited to $<1$~nA. A constant voltage is maintained on a doped Si back gate in order to have consistent contact quality. Data in the inset of Fig.~\ref{fig:2}a and Extended Data Fig.~\ref{fig:4LTransport} are taken in a dilution refrigerator at a nominal base temperature of 20 mK, at zero magnetic field unless otherwise specified. Data in Fig.~\ref{fig:2}f are measured in the nSOT microscope with a nSOT held 300 nm above the surface of the device, at ($B_\perp, B_\parallel$) = (0.5~mT, 40~mT) to match the experimental conditions of Fig.~\ref{fig:2}b,e--g.

Transport measurements in Fig.~\ref{fig:3}a were taken at 17.77~Hz under a 3~nA current bias in the pseudo $R_{xy}$ configuration shown in the inset. These measurements were taken under  ($B_\perp, B_\parallel$) = (19~mT, 0~mT).


\subsection*{Square and arbitrary wave pulse modulated magnetometry}

Quasi-d.c. magnetometry measurements in the main text, indicated as $\Delta B$, were primarily acquired by pulse modulating the gates in a square waveform and acquiring data on a lock-in amplifier (either Stanford Research Systems SR830 or Stanford Research Systems SR860) as detailed in Ref.~\cite{patterson_superconductivity_2025}. Most measurements of $\Delta B$ (including all of main text Fig.~1, Fig.~2, and Fig.~4) used Analog Devices analog switches ADG1534 or ADG2436 to square wave modulate between gate voltages of a magnetic state and a reference state at a set measurement frequency. This produces parasitic electric field contrast which is linear in gate voltage and subtracted out. These measurements should be understood as the average difference between the point and the reference, for example in Fig~\ref{fig:3}d each point is referenced to either the low-$n_e$ endpoint (blue curve) or the high-$n_e$ endpoint (red curve), but each pixel is the average of many sweeps between the point and the reference. Some states could not be stabilized in this way as many different domain configurations could form under the tip which would wash out the signal when averaged. To stabilize, we slowly ramp the gate voltage from the reference to the point over some time constant $\tau$, using a custom made 20-bit digital-analog-converters to generate arbitrary waveforms. As shown in Extended data Fig.~\ref{fig:dynamics}, a slow enough ramp has a high probability of stabilizing a single state under the same sweep. Fig~\ref{fig:3}e was measured with $\tau = $~1472~$\mu$s to stabilize the entire device in the negative domain.

\subsection*{Orbital and spin magnetic moments in the presence of in-plane magnetic fields}
As in Ref.~\cite{patterson_superconductivity_2025}, to ensure tip sensitivity, we apply a large in-plane magnetic field of $B_{\parallel} \approx 50$~mT in many measurements in this paper. It is well known that in a spin-polarized, valley-balanced phase (such as a spin-polarized half-metal) and in the presence of only intrinsic spin-orbit coupling, the spin magnetic moment will align with even a relatively small ($\approx 10$~mT) applied magnetic field~\cite{arp_intervalley_2024, auerbach_isospin_2025, patterson_superconductivity_2025}. In a spin- and valley-polarized phase (such as the orbital ferromagnets studied in this manuscript), the effect of intrinsic spin-orbit coupling is much stronger, pinning spin moments out-of-plane even in the presence of dominant in-plane fields up to over $100$~mT~\cite{arp_intervalley_2024, auerbach_isospin_2025}. In the absence of $C_3$ symmetry breaking, (net) orbital magnetic moments are also always oriented out-of-plane. Because the strength of this spin-orbit coupling is independent of the orbital magnetic moment, spin moments remain pinned out-of-plane, even in very low orbital magnetic moment states where spin magnetization dominates. Simulations and images in Fig.~\ref{fig:4} show the fringe field patterns expected for in- and out-of-plane magnetic moments. Conventionally, orbital ferromagnetism has been identified unambiguously by the presence of magnetic moments that exceed the Bohr magneton per charge carrier. However, the effect of spin-orbit coupling can also help us to identify regions of the phase diagram that develop orbital magnetism; even when the total magnetic moment is within experimental uncertainty of a Bohr magneton per carrier, if magnetic moments show a characteristic easy-axis anisotropy, there must be a valley imbalance present. In our measurements of fringe field over the $n_e, D$ tuned region of superconductivity in Fig.~\ref{fig:2} and Fig.~\ref{fig:4}, we make use of this fact to unambiguously demonstrate the presence of orbital magnetism in the superconductor.

\subsection*{Magnetization fit and simulated fringe fields} 
The connection between the underlying sample magnetization and measured fringe magnetic field at a given height depends strongly on geometry. In order to address this, we fit the sample magnetization using a least squares fit to the experimental data. In short, we numerically implement a forward problem which takes in a spatially varying magnetization in a given direction and calculates the magnetic field projected along the direction of the nSOT at the measurement height. We then iteratively minimize the residual between the simulated magnetic field and the measured data, with an additional constraint that penalizes gradients of the magnetization to favor smooth solutions. Fig.~\ref{fig:2}c-d shows one such solution of this algorithm, where the simulated magnetic field (Fig.~\ref{fig:2}c) from the magnetization shown in Fig.~\ref{fig:2}d matches the directly measured fringe field in Fig.~\ref{fig:2}b. One important aspect of this fitting is that it makes clear the behavior of the magnetic fringe field above the nearly uniformly magnetized device: for example, negative fringe field outside the device area from magnetic field lines wrapping around. Further examples are shown in Fig.~\ref{fig:orbital_moment_estimates}. Data in Fig.~\ref{fig:1}c-d are converted from $\Delta B$ to $M$ using the calibration provided from this fitting procedure, including checks that the magnetic domain structure does not clearly change along the path of the measurement. Data in Fig.~\ref{fig:4}f are converted from $\Delta B$ to $M$ similarly, and then is further expressed in $m$, the magnetization per carrier, by dividing by the known carrier density. 

\subsection*{Calibration of the magnetic field in measurements of Device 4L}

For measurements of the superconducting state in Device 4L, it is important to carefully calibrate $B_\perp$ such that the direction of the small, applied magnetic field is known, to disentangle possible signals of orbital ferromagnetism rather than weak diamagnetism. Both trapped flux in the superconducting magnet and sample tilt can cause a few mT difference between the applied $B_z$ on the superconducting magnet power supply and the effective $B_\perp$ felt by the sample \cite{zhang_imaging_2026}. To account for this, we calibrate $B_z$ by measuring the magnetic field at which the known quarter metal (ring of fire) orbital ferromagnet reverses magnetization as a function of applied magnetic field, which we calibrate within 50 $\mu$T. $B_\perp$ as noted in the main text takes into account the measured offset. 

\clearpage
\newpage
\pagebreak

\onecolumngrid

\setcounter{figure}{0}
\renewcommand{\thefigure}{E\arabic{figure}}

\begin{figure}
    \centering
    \includegraphics{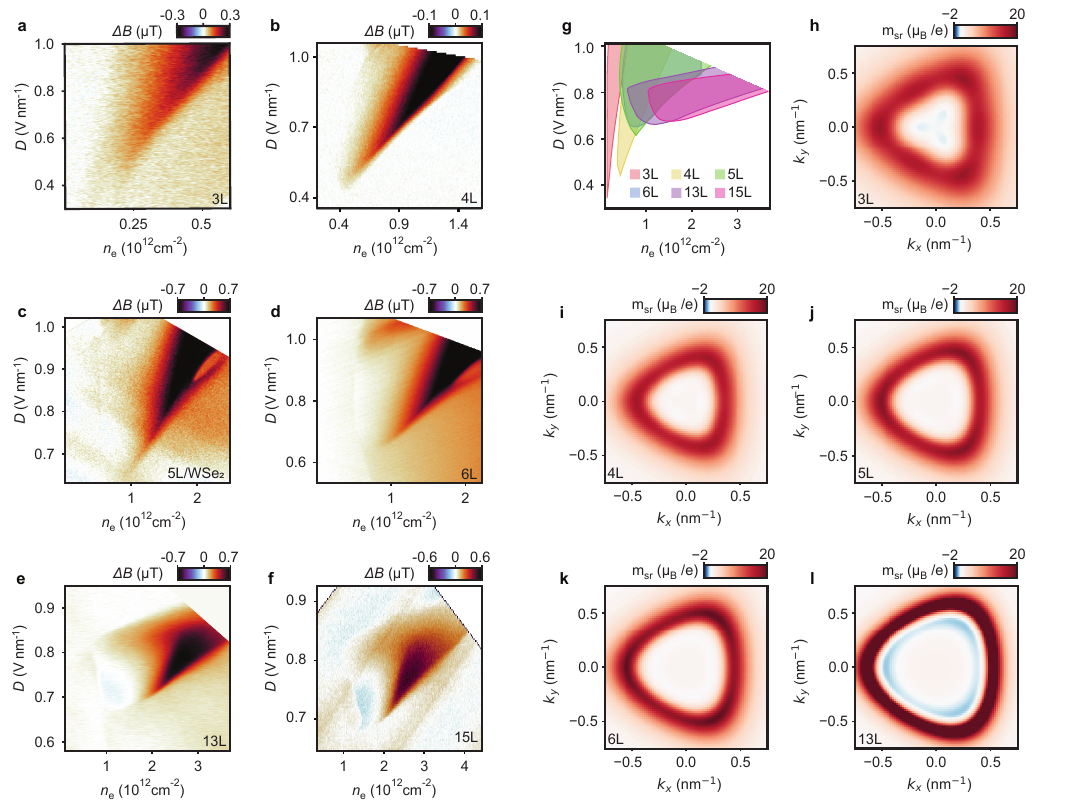}
    \caption{\textbf{Orbital ferromagnetism in 3, 4, 5, 6, 13, and 15 layer rhombohedral graphene} 
    \textbf{(a)}-\textbf{(f)} Phase diagrams of $\Delta B$ versus $n_e$ and $D$ referenced to the spin-polarized, valley-unpolarized half metal state, similar to main text Fig.\ref{fig:1}a,b for all layer numbers considered in this work.
    \textbf{(g)} Cartoon schematic of the rough extent of orbital ferromagnetism for all considered layer numbers on the same $n_e$ and $D$.
    \textbf{(h)}-\textbf{(l)} Calculated self-rotation magnetization as a function of $k_x$ and $k_y$ under the same displacement field, similar to main text Fig.\ref{fig:1}g, for most of the considered layer numbers showing the generality of the so-called Berry ring of fire.}
    \label{fig:all_orbital_m}
\end{figure}

\begin{figure}
    \centering
    \includegraphics{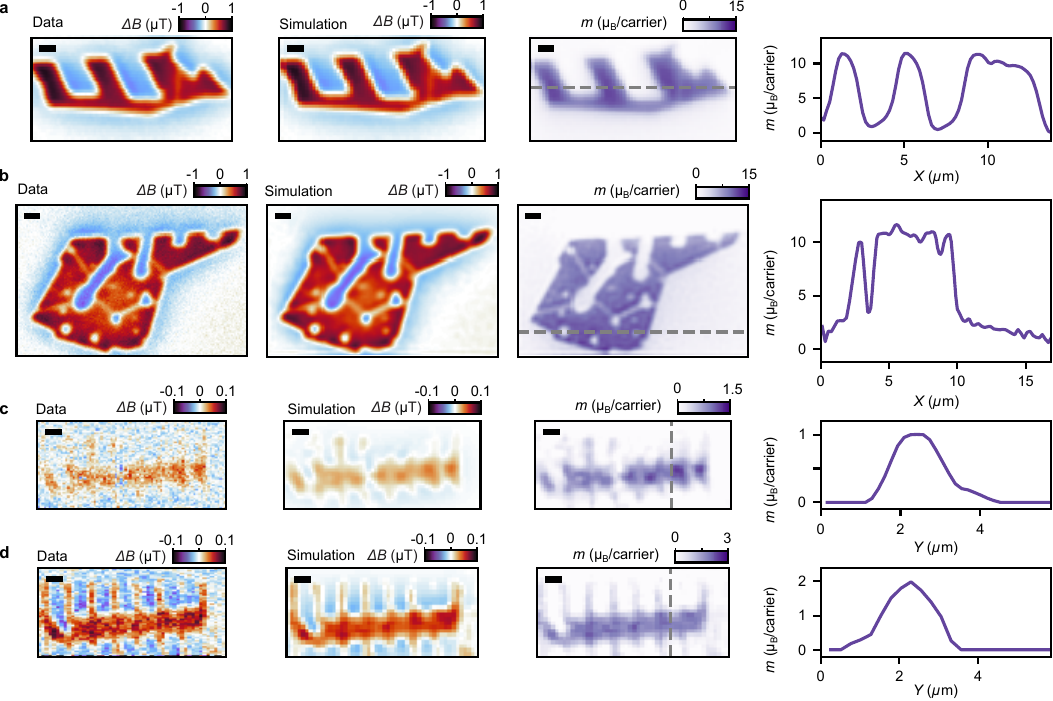}
    \caption{\textbf{Orbital magnetization in Devices 6L, 13L, and 4L.}  
    Summary of fits to the orbital magnetization in the `ring-of-fire' portion of the quarter metal in Device 6L (\textbf{a}) and 13L (\textbf{b}) and the superconducting state in Device 4L (\textbf{c}-\textbf{d}). For each panel, from left to right, we show the measured $\Delta B$ signal at a finite height 200-350 nm above the rhombohedral graphene (first column), the simulated $\Delta B$ (second column) corresponding to the best fit magnetization per carrier $m$ (third column), and $m$ as a function of spatial position along the dashed trajectory (fourth column). Data in panel \textbf{a} is taken at $n_e = 2.44$ $10^{12}$ cm$^{-2}$, $D = 0.76$ V nm$^{-1}$ with a reference point of $n_e = -0.058$ $10^{12}$ cm$^{-2}$, $D = 0.71$ V nm$^{-1}$. Data in panel \textbf{b} is the same as in Fig.~\ref{fig:2}f. Data in panel \textbf{c} is the same as in Fig.~\ref{fig:2}e. Data in panel \textbf{d} is the same as in Fig.~\ref{fig:4}c. All reference points for panels \textbf{b-d} are specified in the Methods. All scale bars are 1 $\mu$m. 
    }
    
    \label{fig:orbital_moment_estimates}
\end{figure}

\begin{figure}
    \centering
    \includegraphics{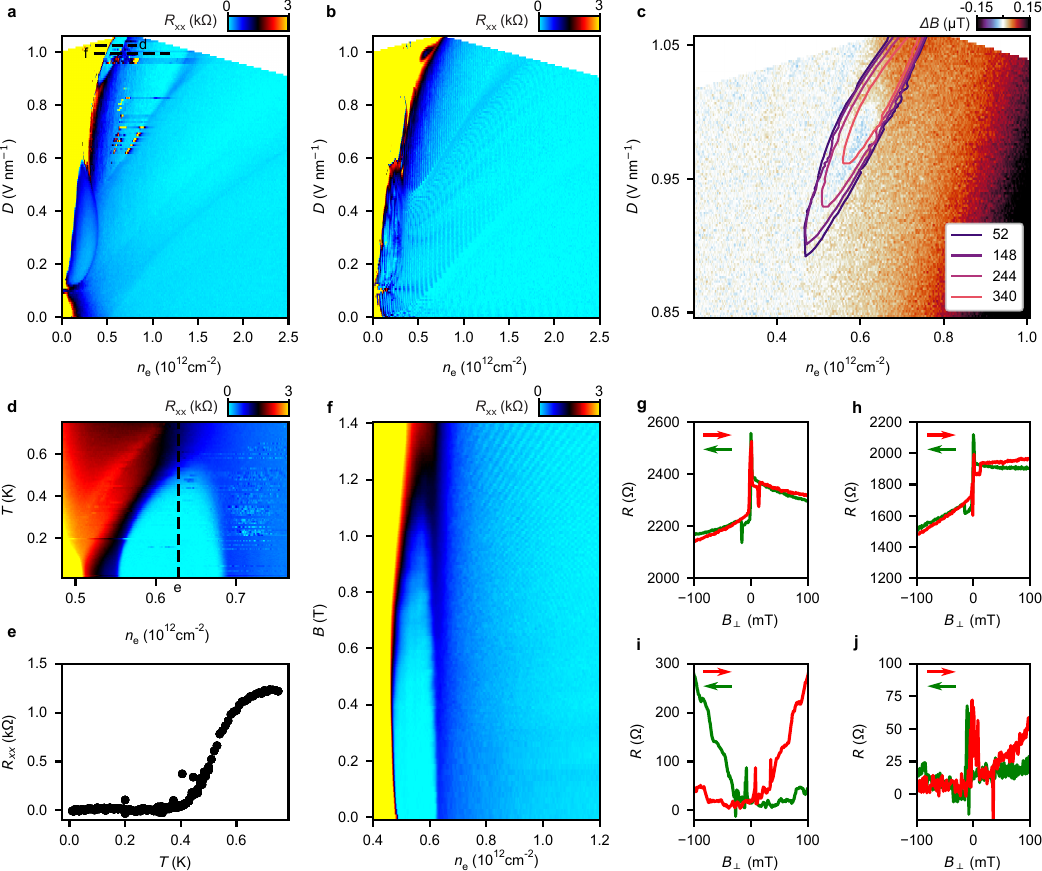}
    \caption{\textbf{Electronic transport characterization of device 4L} 
    \textbf{(a)} $R_{xx}$ as a function of $n_e$ and $D$ of the electron doped bands in device 4L. Dashed lines indicate linetraces plotted in panels d,f. Data is taken at $T = 20$ mK and $B = 0$.
    \textbf{(b)} $R_{xx}$ as a function of $n_e$ and $D$ measured at $T = 20$ mK and $B = 1$ T, showing singly, doubly, and fourfold degenerate quantum oscillations in the quarter, half, and unpolarized metals respectively. 
    \textbf{(c)} $\Delta B$ data shown in Fig. \ref{fig:4}a in the main text, overlaid with contours of constant resistance (approximately 20\% of the normal state resistance) at various temperatures. 
    \textbf{(d)} $R_{xx}$ as a function of $n_e$ and $T$ across the chiral superconductor. 
    \textbf{(e)} $R_{xx}$ as a function of $T$ at the approximate center of the chiral superconducting region of parameter space, marked by dashed line in panel d. The phase has a $T_c$ around $400$ mK. 
    \textbf{(f)} $R_{xx}$ as a function of $n_e$ and $B_\perp$ around the chiral superconductor. The superconducting phase persists to relatively high $B_\perp$, and adjacent (singly degenerate) quantum oscillations confirm the nearby phase to be a quarter metal. 
    \textbf{(g-j)} Measurements of the resistance in longitudinal (\textbf{g},\textbf{i}) and nonlocal, Hall-like contact configurations (\textbf{h},\textbf{j}). These data are measured at a fixed gate voltage in the center of the magnetic superconducting state, at $(n_e, D) = (0.59\times 10^{12} \textrm{cm}^{-2},1 \textrm{V/nm})$. Panels \textbf{g} and \textbf{h} are measured at $T = 500$ mK (above $T_c$) while \textbf{i} and \textbf{j} are measured at $T = 28$ mK (below $T_c$). Curves are measured with $B_\perp$ swept in both directions, up to $B_\perp = 300$ mT, where red is sweeping from negative to positive $B_\perp$ and green is positive to negative.
    }
    \label{fig:4LTransport}
\end{figure}

\begin{figure}[ht!]
    \centering
    \includegraphics[width=\columnwidth]{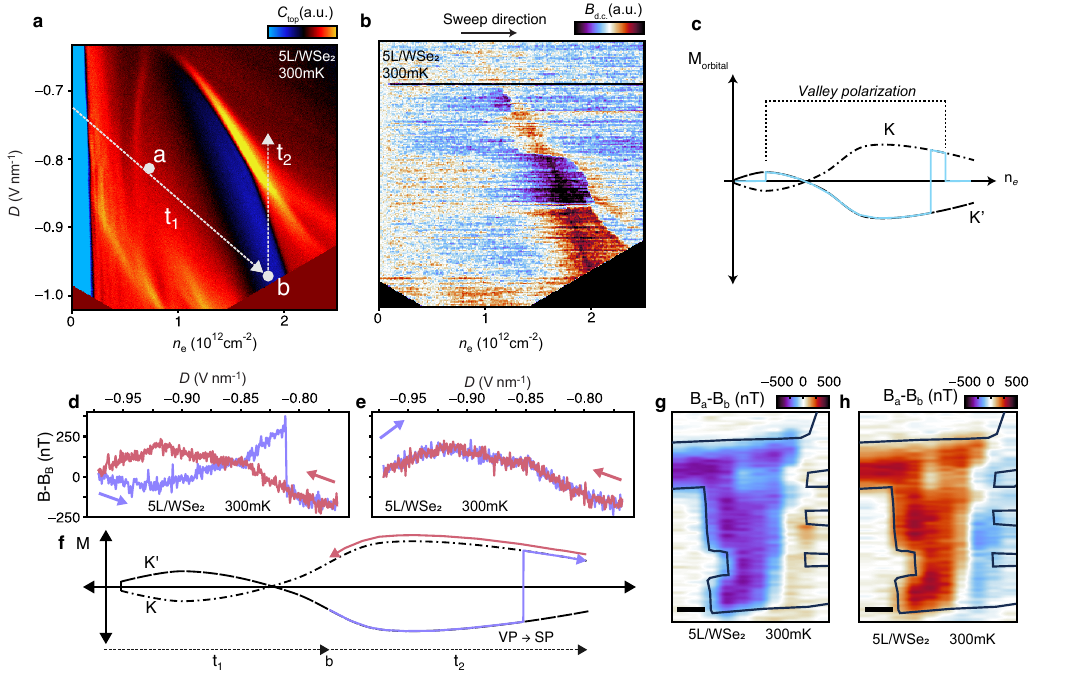} 
    \caption{\textbf{Orbital switching in a 5 layer device}
    \textbf{(a)} Capacitance phase diagram of the electron side quarter metal of a 5-layer RNG device with WSe$_2$ on one side, in the negative $D$ regime that layer-polarizes carriers away from the  WSe$_2$. 
    \textbf{(b)} A DC nSOT magnetic phase diagram of the quarter metal regime, demonstrating that it is possible to observe the orbital switching directly, since in the high density regime the RNG magnetism is large enough to overcome the 1/f noise that motivates differential measurements.
    \textbf{(c)} Schematic of orbital magnetism versus in either valley (dashed lines) and the what might result for a given expected gate trajectory (blue line).
    \textbf{(d)} Valley hysteresis in gate trajectory. State initialized in K' (negative) valley along trajectory t$_1$ in (a), then measuring along t$_2$ while square wave referenced to point b. Near $D = -0.81$~V/nm the state crosses the boundary out of the quarter metal phase and re-enters from the spin-polarized valley-unpolarized state into valley K, resulting in a large switch in the observed magnetism. Sweeping back along t$_2$ shows opposite magnetism (blue line), and a repeated sweep (without re-initialization) in \textbf{(e)} shows only single valley K response. 
    \textbf{(f)} Summarizes the valley state and resulting $M$ as the system evolves along t$_1$ and t$_2$, (e) corresponds only to a trajectory along the K valley line (dot-dash).
    \textbf{(g)} Spatial image of fringe field while square wave modulating between points a and b, after initialization in the K' valley along t$_1$, showing negative valley domain over most of the device extent (black outline).
    \textbf{(h)} Subsequent repeated image after moving the state along trajectory t$_2$ and switching into valley K, showing an equal and opposite domain to (g).
    Measurements were performed with $(B_{\perp}, B_{\parallel}) = (14.5~\mathrm{mT}, 5~\mathrm{mT})$.
    }
    \label{fig:5L_switching}
\end{figure}

\begin{figure}
    \centering
    \includegraphics{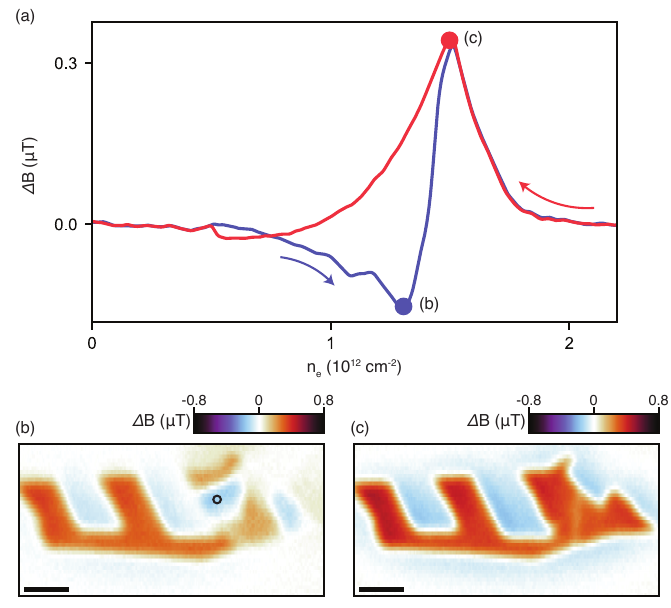}
    \caption{\textbf{Electrically tunable magnetic switching in R6G} 
    \textbf{(a)} Magnetic fringe field, B, versus electron density, $n_{\mathrm{e}}$ of 6 layer rhombohedral graphene at $D = 0.8$~V/nm, sweeping from the unpolarized state at low density (blue) and from the spin-polarized valley-unpolarized state at high density (red). nSOT parked at the marked point in (b).
    \textbf{(b)} Spatial image of B at the blue point in (a) sweeping from low density, exhibiting negative domains that do not extend over the entire device. 
    \textbf{(c)} Spatial image of B at the blue point in (a) sweeping from high density.
    Scale bars are 2$\mu$m.
    }
    \label{fig:6L_switching}
\end{figure}

\begin{figure}[ht!]
    \centering
    \includegraphics[width=\columnwidth]{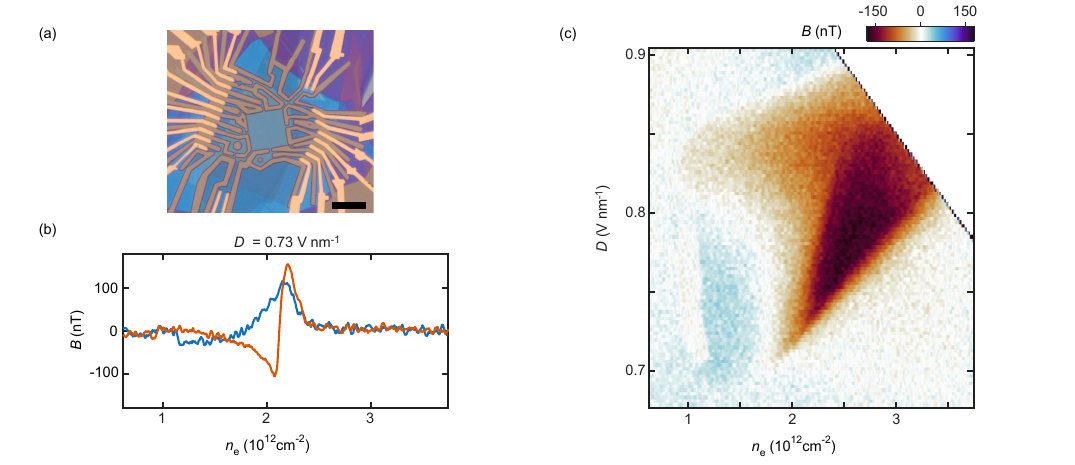} 
    \caption{\textbf{Switching in a second 13 layer device}
    \textbf{(a)} Micrograph of a large area 13-layer RNG device. Scale bar is 20~$\mu$m.
    \textbf{(b)} nSOT gate sweep measuring $\Delta B$ referenced to low and high density, similar to main text Fig.~\ref{fig:3}d, showing orbital switching.
    \textbf{(c)} Phase diagram of $\Delta B$ versus $n_{e}$ and $D$ showing magnetism consistent with main text Fig.~\ref{fig:1}b.
    }
    \label{fig:alt_13L_switching}
\end{figure}

\begin{figure}
    \centering
    \includegraphics{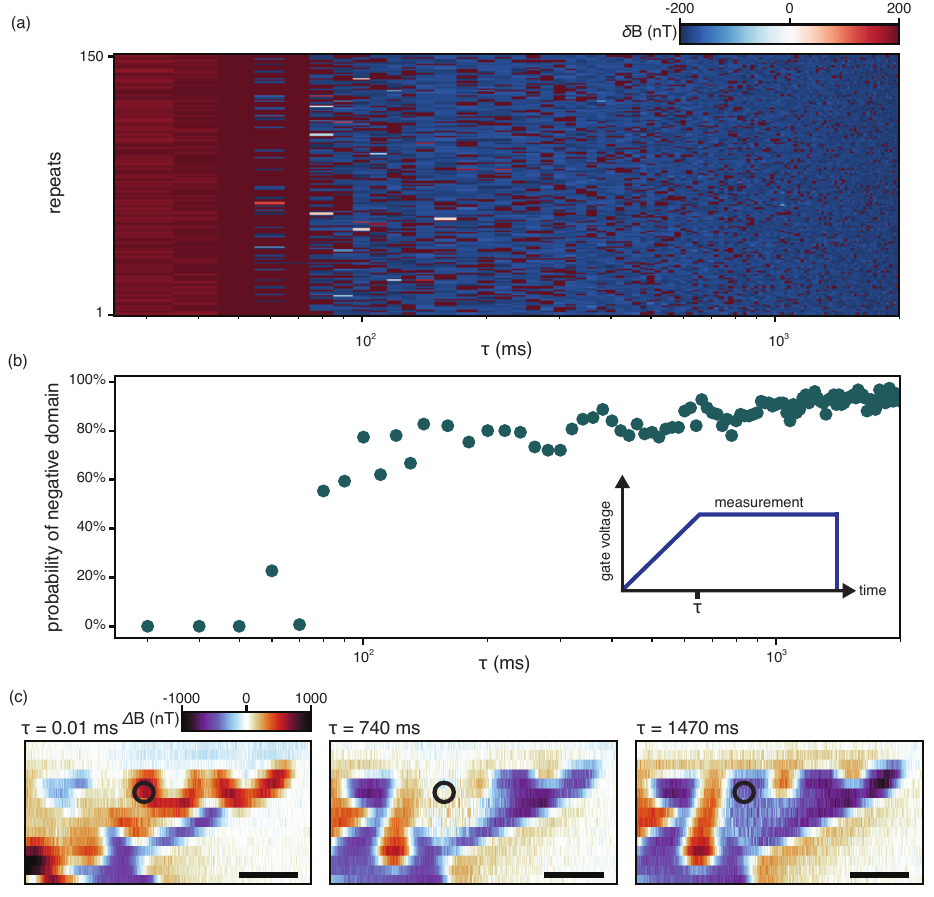}
    \caption{\textbf{Dynamics of negative domain formation} 
    \textbf{(a)}  $\delta B$ measured by repeatedly setting the top and bottom gate voltages over $\tau$ milliseconds, measuring and then zeroing the gate, as shown in the inset of (b). At the location marked in (c).
    \textbf{(b)} Probability of forming a negative domain under the SQUID as a function of rise time $\tau$.
    \textbf{(c)} Spatial measurements of $\Delta B$ for rise times of 0.01~ms, 740~ms, and 1470~ms for left, middle and right respectively. Each pixel is averaged over many gate resets, thus pixels that do not have a high probability of forming a negative or positive domain show weak contrast. Scale bar is 2$\mu$m.
    }
    \label{fig:dynamics}
\end{figure}

\begin{figure}
    \centering
    \includegraphics{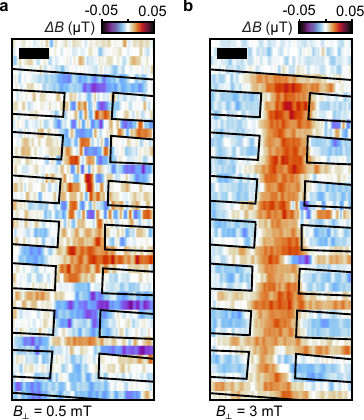}
    \caption{\textbf{Magnetic switching in the normal state of the magnetic superconductor in Device 4L} 
    \textbf{(a)} Real-space imaging of $\Delta B$ for $(B_\parallel,B_\perp) = (40 \text{mT},0.5\text{mT})$ 
    and \textbf{(b)} $(B_\parallel,B_\perp) = (50 \text{mT},2\text{mT})$  
    in the normal state of the magnetic superconductor for Device 4L. Images are taken under identical gate conditions to Fig. \ref{fig:2}e and Fig. \ref{fig:2}g inset, except at elevated temperature ($T \approx 800$ mK). At lower applied $B_\perp$, $\Delta B$ shows numerous line-by-line and within-line `switching' events similar to the single-position measurement of $\Delta B$ in Fig. \ref{fig:4}e. By increasing $B_\perp$ (panel b), switching is suppressed. Scale bars are 1 $\mu$m.
    }
    \label{fig:4L_aboveTc_switching}
\end{figure}

\begin{figure}
    \centering
    \includegraphics{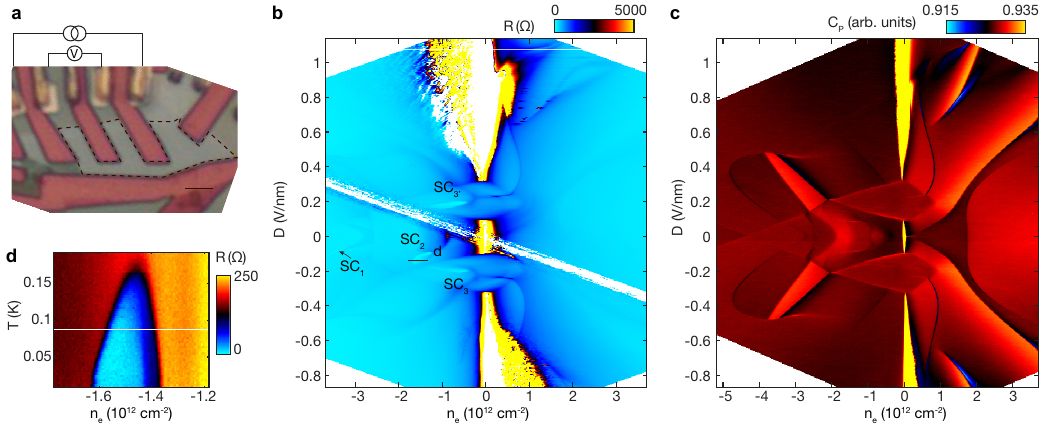}
    \caption{\textbf{Transport in R6G:} 
    \textbf{(a)}  device image and transport measurement configuration of the R6G device. 
    The dashed line outlines the dual-gated rhombohedral region.
    \textbf{(b)} full transport phase diagram as function of charge carrier density $n_e$ and displacement field $D$. 
    Three low $T_c$ superconducting pockets, SC$_{1-3}$ are observed on the hole side.
    Chiral superconducting states on the electron side are not observed.
    \textbf{(c)} capacitance phase diagram over the same density and displacment field range as in panel b.
    \textbf{(d)} example superconducting dome of SC$_2$ at the position marked in panel b.
    }
    \label{fig:hexalayer_transport}
\end{figure}

\begin{figure}
    \centering
    \includegraphics{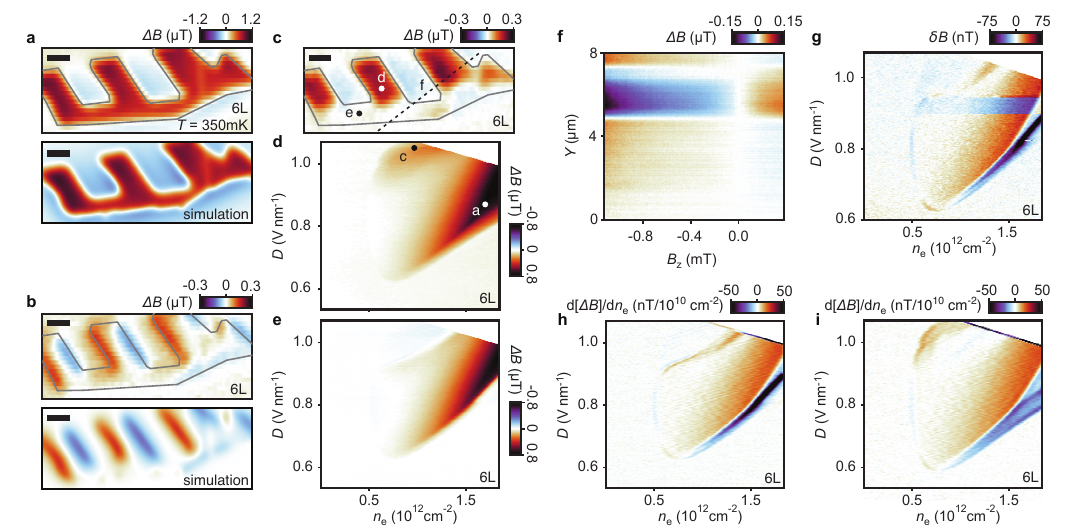}
    \caption{\textbf{Magnetic inhomogeneity in R6G:} 
    \textbf{(a)} Top: Fringe field $\Delta B$ acquired above the 6L device in the quarter-metal phase as indicated on panel d. Bottom: Simulated spatial pattern due to a magnetic moment within the sample geometry ($M = 7 \,\, \mu_B \,\, 10^{12}$ cm$^{-2}$) aligned out-of-plane. 
    \textbf{(b)} Top: Fringe field of a spin-polarized phase at $n_e=-1.77 \times 10^{12}$ cm$^{-2}$, $D = 0$~V~nm$^{-1}$. Bottom: Simulated spatial pattern due to the magnetic moment within the sample geometry in (a) ($M = 1.25 \,\, \mu_B \,\, 10^{12}$ cm$^{-2}$) aligned in-plane in the same direction as the applied field.
    \textbf{(c)} Fringe field at roughly the same density and displacement field where chiral superconductivity has been reported indicated by the point in panel d showing domain structure inconsistent with panel a or b.
    In panels a-c, grey outlines indicate the extent of the dual-gated heterostructure.
    \textbf{(d)}, \textbf{(e)} Fringe field $\Delta B$ as a function of $n_{\textrm{e}}$ and $D$ at the points marked in \textbf{(c)}.
    All data was taken under a large applied in-plane field $(B_{\perp}, B_{\parallel}) = (1~\mathrm{mT},45~\mathrm{mT})$ and at $T = 350$~mK. \textbf{(f)} Field dependence of $\Delta B$ along the spatial linecut indicated in \textbf{(c)}. While \textbf{(e)} appeared to show some negative $\Delta B$ in the region of reported chiral superconductivity, \textbf{(f)} proves that signal was the result of non-local fringe field pickup, owing to the nearby positively magnetized contacts.
    To ensure that our measurement protocol is not somehow inadvertently causing the formation of a stochastic magnetic domain which somehow perfectly averages to fringe field of zero over that entire range of $n_e, D$, we complete small sinusoidal differential magnetometry (which is far less affected by stochastic switching as measurements do not involve repeated field coolings) in \textbf{(g)} taken in the region of small or zero magnetization away from the contacts. This measurement can be compared to the numerical derivatives of \textbf{(d)} and \textbf{(e)} shown in \textbf{(i)} and \textbf{(h)}, respectively. \textbf{(g)} shows broad agreement with \textbf{(h)} and not with \textbf{(i)}, substantiating that the results of \textbf{(d)} and \textbf{(e)} are not originating from a fine tuned stochastic magnetic switching effect.}
    \label{fig:hexalayer_inhomoheneity}
\end{figure}

\clearpage
\newpage
\pagebreak

\onecolumngrid

\setcounter{equation}{0}
\setcounter{figure}{0}
\setcounter{table}{0}
\setcounter{page}{1}
\setcounter{section}{0}
\makeatletter
\renewcommand{\theequation}{S\arabic{equation}}
\renewcommand{\thefigure}{S\arabic{figure}}
\renewcommand{\thepage}{\arabic{page}}

\clearpage
\newpage
\pagebreak

\onecolumngrid

\begin{center}
\textbf{\large Supplementary information for `\papertitle'}\\[5pt]
\end{center}

\bigskip
This supplementary information includes the following sections:
\begin{enumerate}
    \item Gate tuned hysteresis of the magnetic moment
    \item Electric field background subtraction
    \item Stability of the superconducting state to nSOT measurements 
    \item Disorder in the region of magnetic superconductivity
    \item Theoretical Calculations
\end{enumerate}

\newpage
\pagebreak

\section{Gate tuned hysteresis of the magnetic moment}
\begin{figure}
    \centering   \includegraphics[width=90mm]{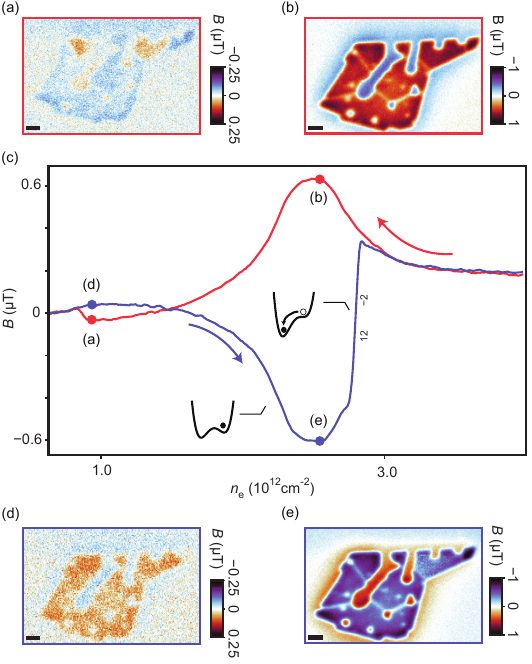} 
    \caption{\textbf{Additional characterization of the magnetic hysteresis in Fig.~\ref{fig:3}.}
    \textbf{(a)}, \textbf{(b)}, \textbf{(d)}, and \textbf{(e)} Images corresponding to the points indicated in the linecut in \textbf{(c)}. The data of \textbf{(b)}, \textbf{(c)}, and \textbf{(e)} is identical to that in the main text. Here we additionally present data showing that the sign switching we observe at low density is present across the device in \textbf{(a)} and \textbf{(d)}. 
    }
    \label{fig:supp_switching}
\end{figure}

\newpage
\pagebreak

\section{Electric field background subtraction}

Our primary form of parasitic contrast comes in the form of electric field sensitivity. This contrast is most sensitive to changes in the voltage of the top gate as it is unscreened. All measurements reported here are completed with nanoSQUIDs exhibiting electric field sensitivity qualitatively similar to that of Ref.~\cite{patterson_superconductivity_2025}, that is to say with a d.c. sensitivity which is highly linear in applied electric field (unlike that of Ref.~\cite{tschirhart_intrinsic_2023}, for example). As one might expect, the magnitude of the electric field background varies depending on location above the sample. Below we outline the two primary protocols we take to eliminate electric field backgrounds: first how we eliminate background signals in our $n_e, D$ dependent phase diagrams, and next how we eliminate them for spatial images. Notably, Fig.~\ref{fig:background4L} is not the only way we avoid electric field backgrounds in spatial images, when possible we also simply ensure that the square wave excitation is quite small in units of voltage (in other words, the reference point and magnetic point which we modulate between are close in gate voltage) such that the electric field contrast is negligible. This is the same protocol as was outlined in Ref.~\cite{patterson_superconductivity_2025}.
\newpage
\pagebreak

\begin{figure}
    \centering   \includegraphics{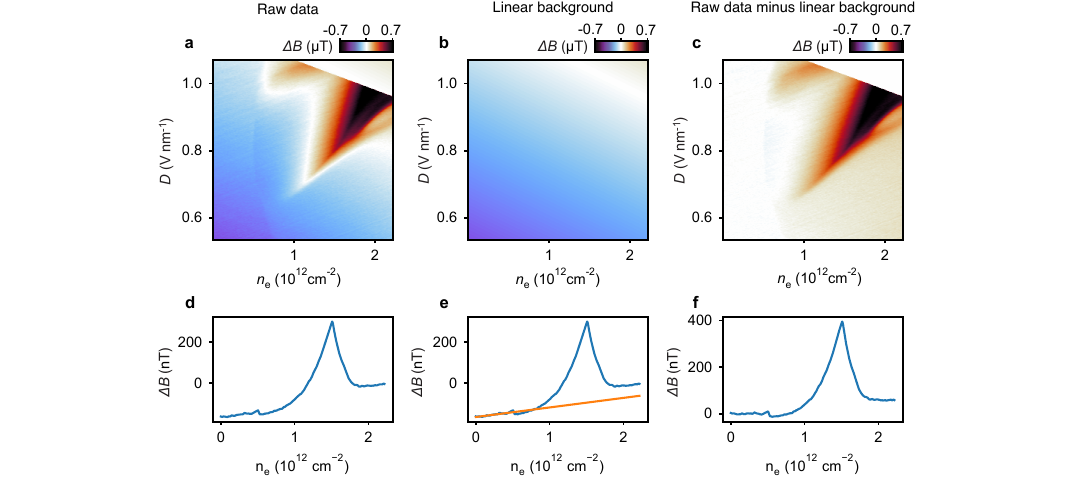}
    \caption{\textbf{Electric field background subtraction in $n_e, D$ phase diagrams.} (\textbf{a}) Typical raw $\Delta B$ signal. Notably an overall linear electric field background can be seen present even in the single particle gap. We fit a small range of the data at low density to a plane to generate the electric field background visualized in (\textbf{b}). The result of subtracting this background from (\textbf{a}) is shown in (\textbf{c}). (\textbf{d}) Similar to (\textbf{a}), but for a typical single line of data. Here we subtract a line fit to the very low density portion of the curve (shown in (\textbf{e})) and subtract it away, shown in (\textbf{f}). These data are the same as in Fig.~\ref{fig:1}a and c, and the conditions under which they were acquired can be found in the Methods section of the main text.}
    \label{fig:backgroundPD}
\end{figure}

\begin{figure}
    \centering   \includegraphics{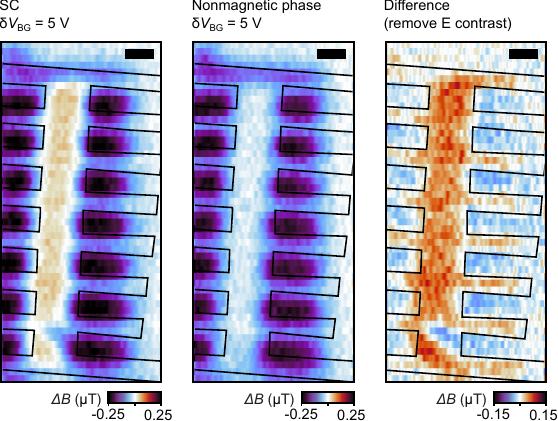}
    \caption{\textbf{Electric field background subtraction in spatial images.} (\textbf{a}) Spatially resolved $\Delta B$ for $n_e=0.71 \times 10^{12}$ cm$^{-2}$, $D = 1.05$~V~nm$^{-1}$ and $n_e^\text{ref}=3.96 \times 10^{12}$ cm$^{-2}$, $D^\text{ref} = 0.76$~V~nm$^{-1}$. This image mixes parasitic electric field background with the magnetic field contrast.
    (\textbf{b}) Spatially resolved $\Delta B$ for $n_e=0 \times 10^{12}$ cm$^{-2}$, $D = 0.875$~V~nm$^{-1}$ and $n_e^\text{ref}= 3.25\times 10^{12}$ cm$^{-2}$, $D^\text{ref} = 0.58$~V~nm$^{-1}$. This represents the same gate step size away from any magnetic features in the phase diagram. As such, it can be interpreted as purely electric field contrast. Subtracting the two signals yields the magnetic contrast which we report, shown in  (\textbf{c})}
    \label{fig:background4L}
\end{figure}

\newpage
\pagebreak

\section{Stability of the superconducting state to nSOT measurements}

\begin{figure}[h!]
    \centering   \includegraphics{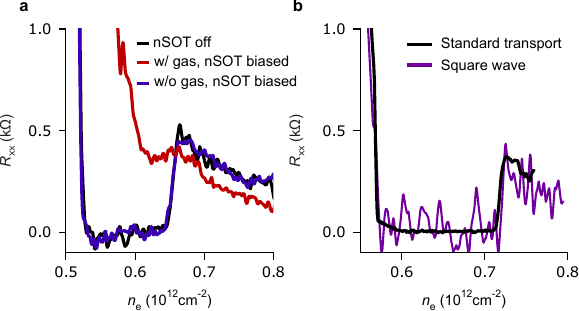}
    \caption{\textbf{Characterizing extrinsic heating in Device 4L.}
    \textbf{(a)} $R_{xx}$ in device 4L as a function of $n_e$ at fixed $D = 0.98$ V/nm in the nSOT microscope under distinct heating conditions. In black, the nSOT is grounded. In red, the nSOT is biased to the measurement working point, 100s of $\mu$A of current bias, with He exchange gas left in the microscope. In blue, the nSOT is biased to the measurement working point but with the sample and tip in vacuum, after pumping out the exchange gas. All measurements are done with $T= 50$ mK at the mixing chamber. 
    \textbf{(b)} $R_{xx}$ in device 4L as a function of $n_e$ at fixed $D = 1.02$ V/nm. Data is measured in two ways: `standard transport' uses the standard low-frequency lockin measurement technique described in the Methods. In purple, the `square wave' transport is measured using a similar square wave modulation of the back gate (as used in nSOT measurements) in addition to a simultaneous square wave of the transport current bias. With increased noise, the two techniques match, showing that the square-wave gate oscillation does not cause sufficient heating to affect the superconducting state.}
    \label{fig:nSOTheating_SW}
\end{figure}

In this section, we expand on the characterization we do of the effect of nSOT measurements on the superconducting state. In Fig. \ref{fig:nSOTheating_SW}a, we present transport data of Device 4L with a nSOT held near the surface with a typical measurement current bias (of order few 100 $\mu$A) through the nSOT. When the sample is in He exchange gas, the superconducting state is completely killed from heating due to the nSOT, to a temperature estimated to $T \approx 800$ mK (significantly above $T_c$). When the sample is in vacuum, there is no observed difference when the nSOT is subjected to a current bias. We also note that there is no observable difference when the nSOT is near the sample versus retracted far away in this condition. 

Beyond static heating from the nSOT, the square-wave gate oscillation used to measure the differential fringe field could, in principle, lead to transient heating of the sample and affect the superconductivity. To ensure this is not the case, we measure transport in the device with a square wave gate oscillation and a simultaneous current oscillation, such that the current is set to 0.5 nA during one half of the square wave duty cycle, and off during the other half of the duty cycle. The voltage probes are then measured using a Zurich Instruments MFLI MF-BOX
Boxcar Averager. While this measurement vastly increases the noise, the measured resistance and width of the superconducting region is not obviously changed, as can be seen in Fig. \ref{fig:nSOTheating_SW}b. 

\newpage
\pagebreak
\section{Disorder in the region of magnetic superconductivity in R6G}

In the 6-layer sample investigated in this manuscript we find evidence of spatial inhomogeneity arising from competing \textit{ground} states, tuned by spatial variation in the underlying electronic Hamiltonian.  Notably, this source of inhomogeneity appears to be particularly pronounced in the domain of $N$, $n_e$, and $D$ where chiral superconductivity is reported in the literature. 
There have been two reports of chiral superconductivity in 6 layer graphene~\cite{qin_extreme_2026, han_evidence_2026}. However, the R6G sample studied here shows no signatures in transport of superconductivity in the electron-doped quarter metal, despite the observation of several superconductors for hole doping with superconducting $T_c\approx 100\text{ mK}$ considerably lower than expected for chiral superconductors (see Fig.~\ref{fig:hexalayer_transport}).

Our measurements outlined in Fig.~\ref{fig:hexalayer_inhomoheneity} broadly show uniform magnetization everywhere except the high-$D$ low-$n_e$ corner of orbital magnetism. In this corner measurement of $\Delta B$ at fixed points in space as well as spatial maps at fixed $n_e, D$ show clear signatures of two regions of the device with independent ground states.  
We have checked with alternative measurement techniques ($\delta B$) as well as with varying out of plane field to ensure that these measurements are not originating from fine tuned or pinned magnetic domains.
It is difficult to resolve these findings with the typical nature of ferromagnetic disorder in rhombohedral multilayers. Typically, regions of the device are `dirty' (e.g. from the presence of polymer residue) and show no magnetism at all, however different ground states in different regions of the device have not been observed before as far as we are aware. 
Few explanations can easily make sense of these results because the effect is quite subtle; magnetism for all other densities and displacement fields which we checked is nominally uniform, only within the high-$D$ low-$n_e$ corner of orbital magnetism does the lack of uniformity become apparent. 
We speculate that this disorder may be coming from some type of strain gradient across the sample, perhaps most concentrated closer to the contacts where metallic electrodes could strain the graphite slightly. 
nanoSQUID-on-tip measurements in bernal trilayer graphene have used de Haas van Alphen quantum oscillations to probe spatially resolved local strain~\cite{zhou_imaging_2023}, however to the best of our knowledge, there have been no findings of spatially varying ground state as a result of this strain.

Comparing the 6L device (which does not exhibit chiral superconductivity) with the 4L device that does (Fig.~\ref{fig:2}) suggests the magnetized domains in the contacts in Fig.~\ref{fig:hexalayer_inhomoheneity} represent the superconducting parent state. Practically, the existence of a competing ground state may explain why magnetic superconductivity appears in some samples but not in others, in marked contrast to other correlated electron states in rhombohedral graphene. Strain may couple particularly strongly to nematic phases, which are known to be prevalent at low density and high displacement field~\cite{li_transdimensional_2026}. If either the magnetic parent state or nonmagnetic competing state (or both) are nematic, then small differences in strain can be expected to tilt the outcome of the resulting competition locally.



\newpage
\pagebreak

\FloatBarrier
\section{Theoretical Calculations}
We perform a mean-field calculation to  account for the electrostatic screening in rhombohedral multilayer graphene (RMG). Screening effects become more important for larger layer numbers, and  including screening is necessary to reliably compute the orbital magnetization for thicker RMG systems.

\subsection{Tight-Binding Hamiltonian}

We use the following continuum single-particle Hamiltonian for RMG near the $\pm\bm{K}$ valleys of the atomic Brillouin zone. The intralayer Hamiltonian is:
\begin{equation}
H_{\textrm{RMG}}(\bm{k})_{l,\sigma;l,\sigma'}=
\begin{pmatrix}
0 & v_{0}(\tau k_x+ik_y)\\
v_{0}(\tau k_x-ik_y) & 0
\end{pmatrix}_{\sigma,\sigma'}.
\end{equation}
The interlayer part is:
\begin{equation}
H_{\textrm{RMG}}(\bm{k})_{l,\sigma;l+1,\sigma'}=
\begin{pmatrix}
v_4(\tau k_x-ik_y) & v_3 (\tau k_x+ik_y)\\ 
t_1 & v_4(\tau k_x-ik_y)
\end{pmatrix}_{\sigma,\sigma'},
\end{equation}

\begin{equation}
H_{\textrm{RMG}}(\bm{k})_{l,\sigma;l+2,\sigma'}=
\begin{pmatrix}
0 & t_2\\ 
0 & 0
\end{pmatrix}_{\sigma,\sigma'}.
\end{equation}
Here $l=1,...,N_L$ labels the layer, with $N_L$  the total number of layers. $\sigma$ is the sublattice index, $\tau=\pm 1$ is the valley index. The velocities in the Hamiltonian  are defined by $v_i=\frac{\sqrt{3}a_0}{2} t_i$, where $a_0=0.246 \textrm{nm}$ is the lattice constant of monolayer graphene. The hopping parameters are~\cite{PhysRevB.108.155406}:
\begin{equation}
(t_0,t_1,t_2,t_3,t_4)=(\text{2600, 356.1, -8.3, -293, -144}) \text{meV}.
\end{equation}

\subsection{Mean-Field Calculation}
We model the RMG/top-gate/back-gate system as a parallel-plate capacitor to account for electrostatic screening.  We use $n_{l}$ to denote the layer-resolved carrier density in RMG ($n_l>0$ is electron doping), where $l=1,...,N_L$ is the layer index. Charge neutrality of the full gate–sample–gate system implies the carrier density in the two gates can be denoted as
\begin{equation}
n_{\textrm{TG}}=-\frac{\sum^{N_L}_{l=1}n_l}{2}+\delta,\quad  n_{\textrm{BG}}=-\frac{\sum^{N_L}_{l=1}n_l}{2}-\delta.
\end{equation}
Here, $\delta$ measures the charge density difference of the two gates. In a dual-gated experiment, the top- and the back-gate voltages are independently tunable, making it possible to tune the displacement field $D$ and the total carrier density in the sample $\sum^{N_L}_{l=1}n_l$ independently. These two quantities are fixed parameters in our calculation. We then solve self-consistently for the layer-resolved carrier densities, as explained below.

$\delta$ is related to the displacement field  $D$ (pointing from back gate to top gate) by:
\begin{equation}
D=\delta e. 
\end{equation}
Given the displacement field $D$, as well as the layer-resolved carrier density $n_l$, we can solve for the electrostatic potential $\phi(z)$ at any point between the two gates:
\begin{equation}
\phi(z)=-\int^{z}_{0} E(x)dx,
\end{equation}
\begin{equation}
E(z)=\sum_{z_l<z}\frac{-en_l}{2\epsilon_0 \epsilon_r}+\sum_{z_l >z}\frac{e n_l}{2\epsilon_0 \epsilon_r}+\frac{-en_{\textrm{BG}}}{2\epsilon_0 \epsilon_r}+\frac{e n_{\textrm{TG}}}{2\epsilon_0 \epsilon_r}.
\end{equation}
Here, $E(z)$ is the electric field between the two gates, and $z_l$ is the vertical position of the layer $l$. We place the back gate at $z=0$ and choose $\phi(0)=0$. In our calculation, we take $\epsilon_r=4$. We assume a 30nm separation between each gate and the nearest RMG surface, and take the RMG interlayer spacing to be 0.335nm.

After obtaining the electrostatic potential at each layer of RMG, $\phi(z_l)$, we add the following diagonal Hamiltonian to the RMG Hamiltonian:
\begin{equation}
[H_{\phi}]_{l,\sigma;l,\sigma}=-e\phi(z_l).
\end{equation}
We then diagonalize the total Hamiltonian $H_{\textrm{RMG}}(\bm{k})+H_{\phi}$ to obtain the eigenvalues $\epsilon_n(\bm{k})$  and eigenstates $\ket{u_n(\bm{k})}$. Here $n$ is the band index of the Hamiltonian. The chemical potential $\mu$ is solved by
\begin{equation}
4\sum_{n\in \textrm{valence},\bm{k}} f(\epsilon_n(\bm{k})-\mu)+ \sum_{n\in \textrm{conduction},\bm{k}} f(\epsilon_n(\bm{k})-\mu)-4N_LN_{\bm{k}}=\sum^{N_L}_{l=1} n_l A.
\end{equation}
Here, $f(\epsilon)=(1+\exp(\epsilon/k_BT))^{-1}$ is the Fermi-Dirac function, and we set $k_BT=0.1\textrm{meV}$ in our calculation.  The prefactor 4 reflects the spin-valley degeneracy of the valence bands, while the doped electron carriers are assumed to occupy a single spin-valley flavor. The final term $-4N_LN_{\bm{k}}$ subtracts the filled-valence-band background charge. $N_{\bm{k}}$ is number of $\bm{k}$ points we sample, with A being the corresponding sample area. For the R13G and R6G calculations, we sample 151 $\times$ 151 $\bm{k}$ points in a square with edge length 2.8$\textrm{nm}^{-1}$ centered at the $\bm{K}$ valley.

From $\ket{u_{n}(\bm{k})}$, we can calculate the (updated) carrier density on each layer:
\begin{equation}
n_l=\frac{4}{A}\sum_{n\in \textrm{valence},\bm{k}} f(\epsilon_n(\bm{k})-\mu) \langle u_{n}(\bm{k})|P_l|u_{n}(\bm{k}) \rangle+\frac{1}{A}\sum_{n\in \textrm{conduction},\bm{k}} f(\epsilon_n(\bm{k})-\mu) \langle u_{n}(\bm{k})|P_l|u_{n}(\bm{k}) \rangle-\frac{4}{A}N_{\bm{k}}.
\end{equation}
In the equation above, $P_l$ is the projector onto the $l$-th layer. We have assumed that the background charge density is distributed uniformly across the layers. 

Finally, we use the updated $n_l$ values to begin the next iteration. This procedure is repeated until convergence.

\subsection{Implementing DIIS}
 Direct inversion in the iterative subspace (DIIS) can be used to accelerate convergence of the mean-field calculation. We briefly outline the implementation below. Interested readers are referred to Ref.~\cite{PULAY1980393}.
 
 For the $i$-th iteration, the input is a layer-resolved carrier density $n_{l,i}$. From that, we obtain the electrostatic profile $\phi_i(z)$ and the updated layer-resolved carrier density $n_{l,i+1}$ as output.  $\Delta_{l,i}\equiv n_{l,i+1}-n_{l,i}$ measures the change of carrier density between iterations. Convergence is reached when $\sum_l|\Delta_{l,i}|$ approaches zero. Without DIIS, $n_{l,i+1}$ would be used as the input for the next iteration.

 We define an $(N_{\textrm{DIIS}}+1)\times (N_{\textrm{DIIS}}+1)$ matrix $B$ whose entries are calculated from  results of the past $N_{\textrm{DIIS}}$ iterations,  $\Delta_{l,i}$, $\Delta_{l,i-1}$... $\Delta_{l,i-N_{\textrm{DIIS}}+1}$ :
 \begin{equation}
 \begin{split}
 &B_{a,b}=\sum_{l}\Delta_{l,i+1-a}\Delta_{l,i+1-b},\quad  a,b \in [1,N_{\textrm{DIIS}}]\\
&B_{a,N_{\textrm{DIIS}}+1}=1, \quad a \in [1,N_{\textrm{DIIS}}]\\
&B_{N_{\textrm{DIIS}}+1,b}=1, \quad b \in [1,N_{\textrm{DIIS}}]\\
&B_{N_{\textrm{DIIS}}+1,N_{\textrm{DIIS}}+1}=0.
 \end{split}
 \end{equation}
In practice, $N_{\textrm{DIIS}}=5$ is sufficient for convergence. Instead of using $n_{l,i+1}$, we use $\tilde{n}_{l,i+1}$ as the input for the next iteration:
\begin{equation}
\tilde{n}_{l,i+1}\equiv \sum^{N_{\textrm{DIIS}}}_{a=1} c_a (n_{l,i+1-a}+\Delta_{l,i+1-a}),\quad c_a=(B^{-1})_{a,N_{\textrm{DIIS}}+1}.
\end{equation}

\subsection{Orbital Magnetization}

After the mean-field calculation has converged, we use the converged electrostatic potential profile $\phi(z_l)$ to calculate the total orbital magnetization per unit area $M_{\textrm{orb}}$. The orbital magnetization can be decomposed into the self-rotation (sr) contribution and the center-of-mass (cm) contribution:
\begin{equation}
M_{\textrm{orb}}=M_{\textrm{sr}}+M_{\textrm{cm}},
\end{equation}
\begin{equation}
M_{\textrm{sr}}=-\frac{e}{A\hbar}\sum_{n \in \textrm{conduction},\bm{k}} f(\epsilon_n(\bm{k})-\mu)\sum_{m\neq n} \textrm{Im}\left( \frac{\langle u_n(\bm{k})|\partial_{k_x} H_{\textrm{RMG}}(\bm{k})|u_m(\bm{k}) \rangle \langle u_m(\bm{k})|\partial_{k_y} H_{\textrm{RMG}}(\bm{k})|u_n(\bm{k}) \rangle }{\epsilon_n(\bm{k})-\epsilon_m(\bm{k})} \right),
\end{equation}

\begin{equation}
M_{\textrm{cm}}=-\frac{2e}{A\hbar}\sum_{n \in \textrm{conduction},\bm{k}} f(\epsilon_n(\bm{k})-\mu)\sum_{m\neq n} \textrm{Im}\left( \frac{\langle u_n(\bm{k})|\partial_{k_x} H_{\textrm{RMG}}(\bm{k})|u_m(\bm{k}) \rangle \langle u_m(\bm{k})|\partial_{k_y} H_{\textrm{RMG}}(\bm{k})|u_n(\bm{k}) \rangle }{(\epsilon_n(\bm{k})-\epsilon_m(\bm{k}))^2} \right) (\mu-\epsilon_n(\bm{k})).
\end{equation}

In the equations above, we have omitted the derivative acting on $H_{\phi}$, since it does not depend on $\bm{k}$. The summation of $n$ is restricted to the conduction bands because the contribution from the valence band cancels after summing over the spin- and valley-degenerate filled valence bands.

\end{document}